\documentclass[a4paper,twoside,12pt]{article}
\usepackage{amsmath,amssymb}
\usepackage{verbatim}
\usepackage{bbm}
\usepackage{bm}
\usepackage{enumerate}
\usepackage{stmaryrd}
\usepackage{graphicx,color}
\usepackage{indentfirst}
\usepackage{amscd}
\usepackage[all]{xy}
\usepackage{multirow}
\usepackage{mathtools}

\usepackage[table,dvipsnames]{xcolor}

\usepackage{titlesec}
\usepackage{titletoc}

\titleformat*{\section}{\large\bfseries}
\titleformat*{\subsection}{\normalsize\bfseries}

\newcommand{\mostimportant}[1]{{\footnotesize #1\par}}

\DeclareMathOperator{\RE}{Re}
\DeclareMathOperator{\image}{image}
\DeclareMathOperator{\rank}{rank}
\DeclareMathOperator{\Hom}{Hom}

\DeclareMathOperator{\Der}{Der}
\DeclareMathOperator{\diag}{diag}
\DeclareMathOperator{\swap}{swap}
\DeclareMathOperator{\Weyl}{Weyl}

\newcommand{\GLS}[1]{\mathcal{G}(#1)}

\newcommand{\overbar}[1]{\mkern 1mu\overline{\mkern-1mu#1\mkern-1mu}\mkern 1mu}

\newcommand{\cc}[1]{\overbar{#1}}
\newcommand{\eps}{\varepsilon}

\newcommand{\dd}{\mathrm{d}}
\newcommand{\p}{\partial}
\newcommand{\DAGGER}{\ast}

\renewcommand{\epsilon}{USE eps INSTEAD}

\newcommand{\C}{\mathbbm{C}}

\newcommand{\R}{\mathbbm{R}}

\newcommand{\cre}{\texttt{e}}
\newcommand{\ann}{\texttt{i}}

\newcommand{\IWeyl}[1]{\mathcal{I}^{#1}}

\newcommand{\rrr}{m}

\newcommand{\gauge}{\texttt{G}}

\newcommand{\Ein}[1]{\mathcal{E}^{#1}}
\newcommand{\EinG}[1]{\mathcal{E}^{#1}_{\smash{\gauge}}}
\newcommand{\D}[1]{\mathcal{L}^{#1}}
\newcommand{\DG}[1]{\mathcal{L}^{#1}_{\smash{\gauge}}}
\newcommand{\unk}{\gamma}
\newcommand{\FANCY}[1]{\mathcal{X}^{#1}}
\newcommand{\KERN}{{\mathcal{J}}}
\newcommand{\XNN}[1]{\mathcal{N}^{#1}}
\newcommand{\msp}[1]{\mathcal{M}^{#1}}

\newcommand{\DERIV}{\mathcal{D}}

\newcommand{\LD}{\Lang\DERIV}

\newcommand{\DB}[2]{[#1,#2]_{\DERIV}}
\newcommand{\tb}[2]{[#1,#2]_{\LD}}
\newcommand{\tbexp}[1]{e^{\tb{#1}{\,\cdot\,}}}

\newcommand{\wbx}[3]{\llbracket #1,#2\rrbracket_{#3}}

\newcommand{\db}[2]{\wbx{#1}{#2}{}}
\newcommand{\eb}[2]{[#1,#2]}

\newcommand{\TCM}{T^{\ast}\hskip-3ptM}

\newcommand{\LIN}{L}
\newcommand{\J}{J}
\newcommand{\NON}{N}
\newcommand{\DNON}{{D\NON}}
\newcommand{\TD}[1]{\widetilde{#1}}

\newcommand{\VV}{V\cc{V}}
\newcommand{\VVreal}{{(\VV)_{\textmd{real}}}}
\newcommand{\VVpos}{{(\VV)_{\textmd{positive}}}}

\newcommand{\ff}{finite-free}

\newcommand{\sref}[1]{\S \ref{sec:#1}}
\newcommand{\aref}[1]{\S \ref{app:#1}}

\newcommand{\rC}{\mathcal{C}}
\newcommand{\rR}{\mathcal{R}}

\newcommand{\Lang}{\mathcal{A}}

\titlecontents{section}[4.2em]{\addvspace{-4pt}}
{\contentslabel{1.5em}}
{}{\titlerule*[0.3pc]{.}\contentspage\rule{3.5em}{0pt}}

\newcommand{\axaux}[1]{$\bm{\mathrm{(a#1)}}$}
\newcommand{\axR}{\axaux{1}}
\newcommand{\axV}{\axaux{2}}
\newcommand{\axderrank}{\axaux{3}}
\newcommand{\axpos}{\axaux{4}}

\newcommand{\ixaux}[1]{$\bm{\mathrm{(i#1)}}$}
\newcommand{\ixherm}{\ixaux{1}}
\newcommand{\ixker}{\ixaux{2}}
\newcommand{\ixpos}{\ixaux{3}}
\newcommand{\ixhermp}{\ixaux{1'}}
\newcommand{\ixkerp}{\ixaux{2'}}
\newcommand{\ixposp}{\ixaux{3'}}

\newcommand{\oxaux}[1]{$\bm{\mathrm{(o#1)}}$}
\newcommand{\oxsp}{\oxaux{1}}
\newcommand{\oxds}{\oxaux{2}}
\newcommand{\oxff}{\oxaux{3}}
\newcommand{\kk}[2]{$\textmd{#1}_{#2}$}

\newcommand{\BG}[1]{\texttt{B}^{#1}}
\newcommand{\bil}[1]{\texttt{b}^{#1}}
\newcommand{\auxx}[1]{\texttt{a}^{#1}}
\newcommand{\Auxx}[1]{\texttt{A}^{#1}}

\newcommand{\je}{\texttt{j}}
\newcommand{\rac}[1]{\texttt{r}^{#1}}

\renewcommand{\subset}{\subseteq}

\newcommand{\km}{\texttt{k}}

\newcommand{\SIG}{{-}{+}{+}{+}}

\newcommand{\twofootnotes}[3]{\begin{samepage}\mbox{#1\footnotemark\textsuperscript{,}\footnotemark}\addtocounter{footnote}{-2}\stepcounter{footnote}\footnotetext{#2}\stepcounter{footnote}\footnotetext{#3}\end{samepage}}

\begin{document}

\noindent{\bf\Large
The graded Lie algebra of general relativity}\\
\noindent\rule{0pt}{30pt}{\bf Michael Reiterer\footnote{School of Mathematics, IAS Princeton, NJ, USA}, Eugene Trubowitz\footnote{Department of Mathematics, ETH Zurich, Switzerland}}
\vskip 10mm
\noindent {\bf Abstract:}
We construct a graded Lie algebra in which a solution to the vacuum
Einstein equations is any element of degree 1
whose bracket with itself is zero.
Each solution generates a cochain complex,
whose first cohomology is linearized gravity about that solution.
We gauge-fix to get a smaller cochain complex
with the same cohomologies
(deformation retraction).
The new complex is much smaller,
it consists of the solution spaces of linear homogeneous wave equations
(symmetric hyperbolic equations).
The algorithm that produces
these gauges and wave equations
is both for linearized gravity and the full Einstein equations.
The gauge groupoid is the groupoid of rank 2 complex vector bundles.
\vskip 8mm

\setcounter{tocdepth}{1}
\tableofcontents



\newcommand{\EMPH}[1]{#1} 
\section{Introduction}\label{sec:intro}

\newcommand{\ASP}[1]{\Ein{#1}}
A real graded Lie algebra is a graded vector space
$\ASP{} = \bigoplus_{k} \ASP{k}$ over $\R$
and
an $\R$-bilinear map
$\eb{\,\cdot\,}{\,\cdot\,}: \ASP{k} \times \ASP{\ell} \to \ASP{k+\ell}$
such that
\begin{align*}
\eb{x}{y} & = -(-1)^{|x||y|}\eb{y}{x}\\
0 & =
  (-1)^{|x||z|} \eb{x}{\eb{y}{z}}
+ (-1)^{|y||x|} \eb{y}{\eb{z}{x}}
+ (-1)^{|z||y|} \eb{z}{\eb{x}{y}}
\end{align*}
if $x \in \ASP{|x|}, y \in \ASP{|y|}, z \in \ASP{|z|}$.
The second identity is called graded Jacobi identity.

In this paper we construct a graded Lie algebra
$\Ein{} = \Ein{0} \oplus \Ein{1} \oplus \Ein{2} \oplus \Ein{3} \oplus \Ein{4}$
in which the set of solutions to the vacuum Einstein equations is\footnote{%
This set can be defined in any graded Lie algebra.
See for example \cite{NR}.
}:
\[
\big\{
\;\unk \in \Ein{1}\;\;\big|\;\; \eb{\unk}{\unk} = 0\;
\big\}
\]
The equation $\eb{\unk}{\unk}=0$
is equivalent to the standard Einstein equations
if $\unk \in \Ein{1}$ satisfies a nondegeneracy condition.
The graded Jacobi identity implies
\[\eb{\unk}{\eb{\unk}{\unk}} = 0
\quad\textmd{for all}\quad \unk\in \Ein{1}
\]
which we call Bianchi-Jacobi identity.
We sometimes
refer to
the Lie algebra $\Ein{0}$ as the infinitesimal gauge group, 
and to $\Ein{1}$ as the search space.

\subsection{The graded Lie algebra $\Ein{}$} \label{sec:eohesd}

The construction of $\Ein{}$ is a functor:
\[
\textmd{(gauge groupoid)}
\;\;\;\longrightarrow\;\;\;\textmd{(real graded Lie algebras)}
\]
By definition, the gauge groupoid is the category given \twofootnotes{by}{%
By contrast, the gauge groupoid of the usual metric
metric formalism has 4-dimensional manifolds as objects,
and diffeomorphisms as morphisms.
Note that 
there is a canonical functor 
from this paper's gauge groupoid to the
gauge groupoid of the metric formalism.
}{%
Actually, the construction of $\Ein{}$ remains functorial for
morphisms that restrict to open subsets of the base manifold,
that is, the construction is local.}:
\begin{center}
\begin{tabular}{r@{\quad}l}
{\bf objects:} & rank 2 complex vector bundles, over a 4-dim base manifold\\
{\bf morphisms:} & vector bundle isomorphisms
\end{tabular}
\end{center}
Here different objects can have
different base manifolds.
For simplicity, we always assume that the base manifolds and vector bundles
are topologically trivial.

We denote by $\rR$ and $\rC$
the commutative rings of smooth real and complex
functions on the base manifold; by $V$ the $\rC$-module
of smooth sections of the rank 2 vector bundle;
and by $\cc{V}$ the complex conjugate $\rC$-module\footnote{
Given $V$, one defines $\cc{V}$ as follows:
$V$ and $\cc{V}$
are the same as sets, but the identity
map, denoted $V \to \cc{V}, v \mapsto \cc{v}$,
is required to be $\rC$-antilinear.
}.
From this data we construct the real graded Lie algebra
$(\Ein{},\eb{\,\cdot\,}{\,\cdot\,})$.
It is a quotient:
\[
\Ein{}\;\;=\;\;\D{}/\IWeyl{}
\]
where $(\D{},\db{\,\cdot\,}{\,\cdot\,})$
is a real graded Lie algebra,
and $\IWeyl{}\subset \D{}$ is a graded ideal.
The following two informal summaries are not required to read this introduction:

{\bf The construction of $\D{}$ in \sref{constrD}.}
Let $\Lang$ be the $\rC$-algebra
generated by $V$ and $\cc{V}$,
the infinite direct sum of all mixed tensor products of $V$ and $\cc{V}$.
Let $\DERIV$ be the
$\rC$-module
of $\C$-linear graded derivations on $\Lang$, which is a Lie algebra.
Let\footnote{%
All implicit products are tensor products over $\rC$,
e.g.~$\LD = \Lang \otimes_{\rC}\DERIV$
and
$\VV = V \otimes_{\rC}\cc{V}$.
}
$\D{k} \subset (\VV)^{\otimes k} \DERIV
\subset \LD$
be the $\rR$-submodule of elements that are `real',
and antisymmetric in the $k$ copies of $\VV$.
Only $\D{0}, \D{1}, \D{2}, \D{3}, \D{4}$ are nontrivial,
and $\D{} = \bigoplus_k \D{k}$
has a real graded Lie algebra structure
$\db{\,\cdot\,}{\,\cdot\,}: \D{k} \times \D{\ell} \to \D{k+\ell}$.

{\bf The construction of $\IWeyl{}$ in \sref{spinideal}.}
We define $\rR$-submodules $\IWeyl{k}\subset \D{k}$.
Set $\IWeyl{0} = \IWeyl{1} = 0$.
Let $\IWeyl{2}$
be the unique irreducible representation
of real dimension 10
 (`spin 2 representation') of the Lorentz group
in the subspace of $\D{2}$ of elements
that annihilate the ring $\rC$;
we actually work with a more explicit definition.
For $k \geq 3$ let $\IWeyl{k} \subset \D{k}$
be the submodule spanned by elements
$c \wedge x$ with $x \in \IWeyl{k-1}$ and `real' $c \in \VV$.
We show that $\IWeyl{} = \bigoplus_{k}\IWeyl{k}$
is an ideal,
$\db{\D{}}{\IWeyl{}} \subset \IWeyl{}$.

The bracket on $\Ein{}$ is defined in terms of the bracket on $\D{}$ by
\[
\eb{x \bmod \IWeyl{}}{y \bmod \IWeyl{}} \;\;=\;\; \db{x}{y} \bmod \IWeyl{}
\]
for all $x,y\in \D{}$.


\subsection{Geometry associated to elements of $\Ein{1}$}\label{sec:crccrc}

Rather than a metric, 
naturally associated to each nondegenerate $\unk \in \Ein{1}$
is a conformal Riemann-Cartan geometry.
By definition,
a conformal Riemann-Cartan geometry
on the 4-dim base manifold
is a pair consisting \twofootnotes{of}{%
A conformal metric $[g]$ is an equivalence class of metrics $g$,
two metrics being equivalent if and only if the first is a
positive-valued function times the second.
}{%
An affine connection is a connection on the tangent
(and hence the cotangent) bundle.
}%
\begin{center}
\begin{tabular}{c@{\qquad}l}
$[\,g\,]$ & a conformal metric with signature $\SIG$\\
\rule{0pt}{14pt}$\nabla$ & an affine connection
\end{tabular}%
\end{center}
such
that for each representative $g$ there is a 1-form with
$\nabla g = \textmd{(1-form)}\otimes g$;
this compatibility condition holds for one representative
if and only if it holds for all.
There are no other conditions, in particular $\nabla$ may have torsion.

The set of conformal Riemann-Cartan geometries is a perfectly fine search space
for solutions to the Einstein equations, in the standard sense.
Namely $([g],\nabla)$ is a solution
if and only if three conditions hold:
the torsion of $\nabla$ vanishes;
$\nabla g = 0$ for some representative $g$,
this representative is then unique up to multiplication by
a constant;
and the Ricci curvature vanishes.

There is a canonical surjective map
\begin{align*}
\{\textmd{nondegenerate elements of $\Ein{1}$}\}
\;&\to\;
\{\textmd{conformal Riemann-Cartan geometries}\}\\
\unk\;\;&\mapsto\;([g],\nabla)
\end{align*}
where nondegeneracy is the pointwise nonvanishing of some matrix determinant.
This map is constructed in \aref{rc}.
It takes solutions of the Einstein equations in $\Ein{}$
to solutions of the standard Einstein equations.

Ignoring the nondegeneracy condition,
note that $\Ein{1}$ is a linear space,
whereas the space of conformal Riemann-Cartan geometries is not linear.


\subsection{Perturbative expansions
and linearized gravity}

Suppose $\unk' \in \Ein{1}$
solves the Einstein equations, $\eb{\unk'}{\unk'} = 0$.
Then the Einstein equations for $\unk = \unk' + \unk''$,
written in terms of $\unk''$, are
\[
\dd' \unk'' + \tfrac{1}{2} \eb{\unk''}{\unk''} = 0
\]
where
the $\R$-linear map
 $\dd' = \eb{\unk'}{\,\cdot\,}: \Ein{} \to \Ein{}$
satisfies\footnote{%
$\eb{\unk'}{\eb{\unk'}{\,\cdot\,}} = \tfrac{1}{2}\eb{\eb{\unk'}{\unk'}}{\,\cdot\,}$
by the graded Lie algebra axioms;
 $\eb{\unk'}{\unk'}=0$ by assumption.
} $\dd'\circ \dd' = 0$.

This equation has Maurer-Cartan form,
and is the starting point for
perturbative expansions about the background $\unk'$:
one has to solve a sequence of
linear inhomogeneous problems
$\dd' \unk'' = \rho$, with unknown $\unk''\in \Ein{1}$,
and with inhomogeneity $\rho \in \Ein{2}$.
Note that necessarily $\dd' \rho = 0$.

We now discuss the 
homogeneous problem $\dd' \unk'' = 0$,
which is linearized gravity about $\unk'$.
One quotients out the
gauge redundancies by declaring:
\[
\{\textmd{physically distinct solutions}\}
\;\;=\;\; \mathrlap{\frac{\{\textmd{all solutions}\}}{\{\textmd{trivial gauge solutions}\}}}\phantom{\textmd{trivial gauge solutions}}
\]
The trivial gauge solutions
are those of the form\footnote{%
Similarly, in the metric formalism,
if $g'$ solves the Einstein equations,
then linearized gravity about $g'$ 
admits, as trivial gauge solutions,
the Lie derivative of $g'$ along any vector field.
}
$\unk'' = \dd' \delta$
with $\delta$ any element of
the infinitesimal gauge group $\Ein{0}$, hence
\[
\{\textmd{physically distinct solutions}\}
\;\;=\;\; \mathrlap{\frac{\ker(\dd': \Ein{1}\to \Ein{2})}{\image(\dd': \Ein{0} \to \Ein{1})}}\phantom{\textmd{trivial gauge solutions}}
\]
This is the first cohomology of the cochain \twofootnotes{complex}{%
By definition, the $k$-th cohomology
is $\ker(\dd':\Ein{k}\to \Ein{k+1})/\image(\dd':\Ein{k-1}\to\Ein{k})$.
}{%
In any graded Lie algebra, given an element of degree 1
whose bracket with itself is zero,
one can associate to it a cochain complex in exactly the same way.
See for example \cite{NR}.
}
\begin{alignat}{10}\label{eq:CCE}
0
&\xrightarrow{\;\;\dd'\;\;}\;& \Ein{0}
&\xrightarrow{\;\;\dd'\;\;}\;& \Ein{1}
&\xrightarrow{\;\;\dd'\;\;}\;& \Ein{2} 
&\xrightarrow{\;\;\dd'\;\;}\;& \Ein{3} 
&\xrightarrow{\;\;\dd'\;\;}\;& \Ein{4}
&\xrightarrow{\;\;\dd'\;\;}\;& 0
\end{alignat}
of $\R$-vector spaces.

One expects linearized gravity
to have `wave equation' character,
but mathematically this is not evident from what we have seen so far.
In \eqref{eq:CCE} the wave character of the cohomologies is
hidden behind too many gauge redundancies.

\subsection{Gauge-fixing
and symmetric hyperbolicity}\label{sec:shmap}

Gauge redundancies can be removed or reduced by gauge-fixing.
We gauge-fix by restricting the
unknown to some \twofootnotes{$\rR$-submodule}{%
For elements of $\Ein{1}$,
being in $\EinG{1}$ is equivalent to some $\rrr$ pointwise linear conditions.
Naturally,
$\rrr$ is equal to the rank of the infinitesimal gauge group $\Ein{0}$.
See \sref{tgc}.
}{%
To compare this to gauges for the vector potential $A$
in electromagnetism, note that the
Maxwell equations for $A$ are 2nd order,
but the Einstein equations for $\unk$ are 1st order.}
\[\EinG{1} \;\subset\; \Ein{1}\]
This gauge-fixing makes the Einstein equations
symmetric hyperbolic, see \sref{nonlingr}.
For simplicity, we only discuss linearized gravity in this introduction.

To be systematic,
we consider the more general problem of
gauge-fixing all the spaces in \eqref{eq:CCE}.
We obtain a new complex 
with the same cohomologies,
that consists of the solution spaces of linear homogeneous wave equations.

\newcommand{\SH}[1]{w^{#1}}
In this approach one first chooses a gauge object
$\gauge$. It is a concrete algebraic object,
namely $\gauge$ is any positive definite Hermitian form
on the $\rC$-module $\cc{V}\DERIV/\KERN$ defined in \sref{alginp}.
The algorithm in \sref{algspec}
produces associated $\rR$-submodules
\[\EinG{k} \;\subset\; \Ein{k}\]
such that always $\EinG{0}=\Ein{0}$
and $\EinG{4}=0$.
Define the $\R$-linear maps
\begin{align*}
\SH{k} \;\;:\;\;\EinG{k} & \;\to\; \Ein{k+1}/\EinG{k+1}\\
x & \;\mapsto\; \dd'x\,\, \bmod \EinG{k+1}
\end{align*}
and observe that the cochain complex
\begin{alignat}{10}\label{eq:CCS}
0
&\xrightarrow{\;\;\dd'\;\;}\;& \ker\SH{0}
&\xrightarrow{\;\;\dd'\;\;}\;& \ker\SH{1}
&\xrightarrow{\;\;\dd'\;\;}\;& \ker\SH{2}
&\xrightarrow{\;\;\dd'\;\;}\;& \ker\SH{3}
&\xrightarrow{\;\;\dd'\;\;}\;& \ker\SH{4}
&\xrightarrow{\;\;\dd'\;\;}\;& 0
\end{alignat}
is well-defined,
where $\ker \SH{k} \subset \EinG{k}$
is the kernel of $\SH{k}$.

We now make the purely algebraic remark
that if the $\SH{k}$ are surjective,
then
\eqref{eq:CCE} and \eqref{eq:CCS} have isomorphic cohomologies.
To see this,
pick a right inverse for the $\SH{k}$, that is, an $\R$-linear map
$h: \Ein{k} \to \Ein{k-1}$ for each $k$
such that $\EinG{} \subset \ker h$
and $\image h \subset \EinG{}$
and 
$\image(\mathbbm{1} - \dd' h) \subset \EinG{}$,
with $\EinG{} = \bigoplus_k \EinG{k}$.
Set
$\pi = \mathbbm{1} - \dd' h - h \dd'$,
which maps $\Ein{k} \to \Ein{k}$
for each $k$, and observe that:
\begin{itemize}
\item $\pi$ is a cochain map from \eqref{eq:CCE} to itself,
$\pi \dd' = \dd' \pi$.
\item $\pi$ descends to the cohomology of \eqref{eq:CCE} as the identity map.
\item $\pi$ is a projection, $\pi^2 = \pi$,
with $\image \pi = \bigoplus_k\ker \SH{k}$.
\end{itemize}
It follows that the cohomologies
of \eqref{eq:CCE} and \eqref{eq:CCS} are isomorphic.
The map $\pi$ is said to be a deformation retraction\footnote{%
By definition:
A cochain map $f: A^k \to A^k$ is a deformation retraction (onto its image)
if $f^2 = f$ and
if $\mathbbm{1}-f$ is homotopic to $0$.
A cochain map
$g:A^k \to A^k$ is homotopic to $0$ if
$g = \dd_A h + h \dd_A$
for some linear $h: A^k \to A^{k-1}$,
with $\dd_A: A^k \to A^{k+1}$ the differential.
} 
of the complex \eqref{eq:CCE}
onto \eqref{eq:CCS}.

Here is the main point: The $\SH{k}$ are wave operators,
namely linear symmetric hyperbolic operators\footnote{%
Symmetric hyperbolicity \cite{FrKO}, \cite{taylor}
is arguably the simplest concept of a `wave operator'.
These are first order operators.
Let $\p_{\mu}$ with $\mu=0,1,2,3$
be the partial derivatives on $\R^4$,
and let $\sigma^{\mu}$ be the Pauli matrices.
An example of a linear symmetric hyperbolic operator is
\begin{align*}
\sigma^{\mu}\p_{\mu} & = \begin{pmatrix}
\p_0 + \phantom{i}\p_3 & \p_1 - i\p_2\\
\p_1 + i\p_2 & \p_0-\phantom{i}\p_3
\end{pmatrix}\;\;:\;\; C^{\infty}(\R^4,\C^2) \to C^{\infty}(\R^4,\C^2)
\end{align*}
In this example,
symmetric hyperbolicity comes down to
the fact that the Pauli matrices are Hermitian,
$(\sigma^{\mu})^{\DAGGER} = \sigma^{\mu}$.
To see how this is used,
note that it is exactly what one needs to
check that, if $\Phi \in \ker (\sigma^{\mu}\p_{\mu})$,
then $j^{\mu} = \Phi^{\DAGGER} \sigma^{\mu} \Phi$
satisfies the `local conservation law' $\p_{\mu}j^{\mu}=0$.
},
by design of
the algorithm that produces
the $\EinG{k}$. This is shown in \sref{lingr}. It follows that:
\begin{itemize}
\item The maps $\SH{k}$ are surjective,
by solving linear inhomogeneous wave equations.
Hence \eqref{eq:CCE} and \eqref{eq:CCS} have isomorphic cohomologies.
We do assume here that the causal structure of $\unk'$ is
simple, namely that $\unk'$ is globally hyperbolic.
One can then pick a right-inverse $h$
by requiring that the elements of $\image h$
vanish on some 3-dim spacelike Cauchy hypersurface.
The map $h$ is a `propagator' in physics terminology.
\item The spaces $\ker \SH{k}$ in \eqref{eq:CCS}
are the solution spaces to linear homogeneous wave equations.
In particular, the map that restricts these
solutions to any 3-dim spacelike Cauchy hypersurface
is bijective\footnote{%
\newcommand{\SPX}[1]{\;\;#1\;\;}%
To give an idea of
the degrees of freedom 
that are available, we summarize the ranks of various spaces,
as modules over the ring $\rR$ of real functions.
By definition $\rrr = \rank \Ein{0}$:%
\[
\begin{array}{l|ccccccc|c}
\;\;\;\;k & <0 & \SPX{0} & \SPX{1} & \SPX{2} & \SPX{3} & \SPX{4} & >4 & \textmd{alternating sum}\\
\hline
\rank \D{k}     & 0 & \rrr & 4\rrr & 6\rrr & 4\rrr & \rrr & 0 & 0\\
\rank \IWeyl{k} & 0 & 0 & 0 & 10 & 16 & 6 & 0 & 0\\
\rank \Ein{k}   & 0 & \rrr & 4\rrr & 6\rrr-10 & 4\rrr-16 & \rrr-6 & 0 & 0\\
\rank \EinG{k} & 0 & \rrr & 3\rrr & 3\rrr-10 & \rrr-6 & 0 & 0 & 4
\end{array}
\]\nopagebreak%
The
$\IWeyl{2}$, $\IWeyl{3}$, $\IWeyl{4}$
are irreducible representations
of type $(2,0)$, $(\tfrac{3}{2},\tfrac{1}{2})$, $(1,0)$
of the Lorentz group respectively,
the representation labeled
by the half-integer pair $(i,j)$ having complex dimension $(2i+1)(2j+1)$.
Since $\SH{k}$ is symmetric hyperbolic,
$\rank \EinG{k} = \rank \Ein{k+1} - \rank \EinG{k+1}$.
}.
\end{itemize}

The wave operators $\SH{k}$
respect the causal structure of $\unk'$,
there are no fictitious faster-than-light modes.
An analogous statement holds when the
$\EinG{k}$ are used for the full Einstein equations
in \sref{nonlingr}.


\subsection{The signature $\SIG$}\label{sec:sigfunc}

The construction of $\Ein{}$, outlined in \sref{eohesd},
does not explicitly refer to the signature $\SIG$. 
The signature is however there implicitly,
due to the following well-known and rather nice fact:

Let $U$ be any 2-dimensional complex vector space.
Then the 4-dimensional real subspace 
of $U\otimes_{\C}\cc{U}$ of
elements that are invariant under
the $\C$-antilinear involution
$u_1 \otimes \cc{u_2} \mapsto u_2 \otimes \cc{u_1}$
has a canonical conformal inner product of signature $\SIG$.
This conformal inner product
can be indirectly defined by saying that
$\{u \otimes \cc{u} \mid u\in U\}$
is exactly one half of its
null cone\footnote{%
Directly:
Pick a $\C$-bilinear antisymmetric map
$\eps: U \times U \to \C$, $\eps \neq 0$.
Define a $\C$-bilinear symmetric map
$(U\otimes \cc{U}) \times (U\otimes \cc{U}) \to \C$
by $(u_1\otimes \cc{u_2},u_3\otimes \cc{u_4}) \mapsto -\eps(u_1,u_3)\cc{\eps(u_2,u_4)}$. Its restriction to the 4-dim real subspace of
$U\otimes \cc{U}$
is a real inner product of signature $\SIG$, by direct verification.
Different choices of $\eps$ give the same conformal inner product.}.
By definition, the null cone is the
zero locus of the quadratic form associated to the inner product.

This is a functor from the groupoid of 2-dimensional complex
vector spaces to the groupoid of 4-dimensional real vector spaces
with conformal inner product of signature $\SIG$.
It motivates constructing $\Ein{}$
based on a gauge groupoid of rank 2 complex vector bundles.


\subsection{Relation to other general relativity formalisms}\label{sec:relgr}

The formalism of this paper is closely related to
the frame formalism\footnote{%
Note that `frame formalism' is an umbrella term.
Not all frame formalisms are alike.
For example, in \cite{NP} and \cite{FrH}
the Einstein equations are only quadratically nonlinear,
a structural feature
that is not shared by `all frame formalisms'.
} 
of Newman and Penrose \cite{NP} and Friedrich \cite{FrH}.
To see this,
suppose for simplicity that we are on $\R^4$,
and start with the space of pairs $(E,\Gamma)$ consisting
of:%
\begin{center}
\begin{tabular}{r@{\quad}l}
$(E_i)$ & real vector fields, $E_i = {E_i}^{\mu}\p_{\mu}$\\
\rule{0pt}{14pt}
$({\Gamma_{ij}}^k)$
& real-valued functions subject to\footnotemark~${\Gamma_{ij}}^{n} \eta_{n k}
+{\Gamma_{ik}}^{n} \eta_{n j} =
\tfrac{1}{2}{\Gamma_{in}}^n \eta_{jk}$
\end{tabular}%
\end{center}\nopagebreak
\footnotetext{%
This condition will give conformal Riemann-Cartan geometries.
To get Riemann-Cartan geometries
use the stronger condition ${\Gamma_{ij}}^{n} \eta_{n k}
+{\Gamma_{ik}}^{n} \eta_{nj} = 0$,
which implies ${\Gamma_{in}}^n=0$.
This second condition is used in \cite{NP} and \cite{FrH}.
}
Here $\p_{\mu}$ are the partial derivatives on $\R^4$;
Latin indices $i,j,\ldots$ run over some four-element index set;
and $\eta$ is the diagonal matrix $\diag(-1,1,1,1)$.

We say that a pair $(E,\Gamma)$
is nondegenerate if $E$ is a frame.
Each nondegenerate pair defines a conformal Riemann-Cartan
geometry 
by\footnote{%
It suffices to specify $g$ and $\nabla$ on a frame,
they extend uniquely to a metric and connection.
}:
\[
g(E_i,E_j) = e^{\textmd{(any function)}} \eta_{ij}
\qquad \nabla_{E_i} E_j = 
{\Gamma_{ij}}^nE_n
\]
The compatibility condition 
$\nabla g = \text{(1-form)}\otimes g$
in \sref{crccrc}
 holds automatically.

The vacuum Einstein equations are homogeneously
quadratic if one uses the pair $(E,\Gamma)$ as the unknown. Explicitly
the Einstein equations are\footnote{%
One consequence of these equations is that the 1-form
defined by $\mu(E_i) = \tfrac{1}{2}{\Gamma_{in}}^n$ is closed, $\dd \mu = 0$.
If $\mu = \dd f$, then
the metric defined by $g(E_i,E_j) = e^f \eta_{ij}$
satisfies $\nabla g = 0$.
}:
\newcommand{\EE}[2]{{E_{#1}}^{#2}}
\newcommand{\GG}[2]{{\Gamma_{#1}}^{#2}}
\begin{alignat*}{6}
\EE{i}{}(\EE{j}{\mu})
- \EE{j}{}(\EE{i}{\mu})
& - 
(\GG{ij}{n}-\GG{ji}{n})\EE{n}{\mu} && \;=\; 0\\
\rule{0pt}{16pt}
\EE{i}{}(\GG{jk}{\ell})
-
 \EE{j}{}(\GG{ik}{\ell})
&- 
(\GG{ij}{n}
- \GG{ji}{n})\GG{nk}{\ell} \\
& -
(\GG{ik}{n}\GG{jn}{\ell}
- \GG{jk}{n}\GG{in}{\ell}) && \;=\; 0 \mod \Weyl
\end{alignat*}
Here $\Weyl$ is the set of
$({W_{ijk}}^{\ell})$
of real-valued functions subject to:
symmetrization in $ij$ gives zero;
antisymmetrization in $ijk$ gives zero;
contracting $j$ and $\ell$ gives zero;
 ${W_{ijk}}^{n} \eta_{n \ell}
+{W_{ij\ell}}^{n} \eta_{n k} = 0$.
To be sure,
the definition of $\Weyl$
does not in any way refer to the unknown $(E,\Gamma)$,
and this is an integral
part of the statement that the equations are quadratic.

In the invariant formalism of the present paper,
the pair $(E,\Gamma)$ becomes the single object $\unk$,
the space $\Weyl$ becomes $\IWeyl{2}$,
and the Einstein equations become
$\db{\unk}{\unk} = 0 \bmod \IWeyl{2}$,
which by $\IWeyl{1}=0$ is equivalent to $\eb{\unk}{\unk} = 0$.

The pair $(E,\Gamma)$ is only half of the unknown
in \cite{NP} and \cite{FrH},
because in these papers the authors deliberately
work with a bigger search space\footnote{%
In the notation of this paper,
the Newman-Penrose formalism
for the vacuum Einstein equations is
$\{
(\unk,\Psi) \in \D{1}\times \IWeyl{2}
\,|\, \db{\unk}{\unk} = \Psi
\,\textmd{and}\,\db{\unk}{\Psi} = 0
\}$.
In NP terminology, $\unk$ comprises
tetrad components and spin coefficients,
$\Psi$ the Weyl curvature scalars.
See also \cite{diamond}.
}.
Friedrich made the important observation that one
can exhibit local well-posedness of the Einstein equations
directly in this formalism.
This requires fixing a gauge, but not all gauges will do.
Friedrich \cite{FrH}
found ad-hoc linear gauge conditions
for $E$ and $\Gamma$ that make the Einstein equations 
symmetric hyperbolic.
We used this strategy of Friedrich's in \cite{focus}.
This approach to gauge-fixing
depends critically on using a search space bigger than
 just the set of $(E,\Gamma)$.

With the $(E,\Gamma)$ search space,
the problem of gauge-fixing
and getting symmetric hyperbolic equations
has to be reconsidered:
ad-hoc strategies tend to not work,
and there are reasons to
dispense with ad-hoc strategies.
A solution is provided by
the gauges algorithm in the present paper,
which subsumes \cite{twelve}.

\subsection{Commutative algebra language}

In this introduction we have used the language of differential geometry.
We will actually develop $\Ein{}$ in the
more economical language of commutative algebra.
This is based on the following abstractions:
\begin{center}
\begin{tabular}{r@{\;\;\;$\rightsquigarrow$\;\;\;} l@{\hskip 10mm}}%
set of
functions on a manifold & commutative ring\\
vector field & derivation on ring\\
set of sections of a vector bundle & module over ring
\end{tabular}
\end{center}
The notion of a `graded derivation' is particularly useful.

\addtocontents{toc}{~\emph{The graded Lie algebra $\Ein{} = \D{}/\IWeyl{}$}\par}

\section{The graded Lie algebra $\D{}$}\label{sec:constrD}

\subsection{Axioms}\label{sec:axioms1}

Let $(\rR,V)$ be a pair such that:
\begin{itemize}
\item[\axR] $\rR$ is a commutative ring,
with a distinguished subring
isomorphic to the real numbers, $\R \subset \rR$,
such that their multiplicative units coincide.
\item[\axV]
$V$ is a \ff\footnote{By definition,
an $\rR$- or $\rC$-module is \ff~if and only if it admits a finite
basis, if and only if there exists a module isomorphism to
$\rR^n$ or $\rC^n$ for some
finite $n$. Over a commutative ring,
this holds for at most one integer $n$, called the rank of the module.}~$\rC$-module of rank 2.\\
Here $\rC = \rR \oplus i\rR$ is the complexification of $\rR$.
\item[\axderrank]
The $\rR$-module $\Der(\rR)$ of $\R$-linear derivations $\rR\to \rR$
is \ff\footnote{%
Then the $\rC$-module $\Der(\rC)$ of $\C$-linear derivations $\rC\to\rC$
is \ff~with the same rank,
since $\Der(\rC) = \Der(\rR) \oplus i\Der(\rR)$.
We have
 identified $\Der(\rR)$
with the subset of elements $\alpha \in \Der(\rC)$
for which $\cc{\alpha(f)} = \alpha(\cc{f})$ for all $f \in \rC$,
by $\C$-linear extension.
}.
\end{itemize}
The canonical conjugation on $\rC$
will be denoted $\rC \to \rC, f \mapsto \cc{f}$.

\subsection{Interlude: The geometric setting as a special case}\label{sec:gs}
The axioms in \sref{axioms1} are satisfied
in the following differential geometric,
and topologically trivial setting:
\begin{itemize}
\item $M$ is a manifold diffeomorphic to $\R^{\dim M}$,
not necessarily $\dim M = 4$.
\item
$\rR$ is the ring of the smooth
$\R$-valued functions on $M$,\\
with the constant functions as the distinguished subring $\R \subset \rR$.
\item $V$ is the $\rC$-module of sections
of
a trivial rank 2 complex vector bundle.
\end{itemize}

\subsection{Conjugate module $\cc{V}$}\label{sec:conjugatemodule}
Let $\cc{V}$ be a copy of $V$ as a set.
Denote the identity map by 
\begin{equation*}
V \to \cc{V}\;,\;v \mapsto\cc{v}
\end{equation*}
There is a unique $\rC$-module structure on $\cc{V}$
that makes this map $\rC$-antilinear,
and 
this defines the conjugate $\rC$-module $\cc{V}$.

\subsection{The algebra $\Lang$ and its grading}\label{sec:freealg}

Let $\Lang$ be the free $\rC$-algebra generated by $V$ and $\cc{V}$:
\[
\Lang \;=\;
\textmd{(infinite direct sum of all tensor products of $V$ and $\cc{V}$)}
\]
Explicitly, up to and including all products of length two,
\begin{equation*}
\Lang \;=\;
\rC
\oplus V \oplus \cc{V}
\oplus VV \oplus V\cc{V} \oplus \cc{V}V \oplus \cc{V}\cc{V} \oplus \ldots
\end{equation*}
The tensor product signs $\otimes_{\rC}$ are suppressed:
juxtaposition always means tensor product over $\rC$,
either for modules or for their elements.

The algebra $\Lang$ is graded by 
the free monoid on the two-element set $\{V,\cc{V}\}$.
The direct summands in the definition of $\Lang$
are called homogeneous subspaces.


\subsection{Graded derivations}\label{sec:gradedderivations}

By definition,
a map $\delta: \Lang \to \Lang$ is a graded derivation if
\begin{itemize}
\item $\delta$ is linear over\footnote{%
Beware that a graded derivation may act nontrivially on the ring $\rC$.
} $\C$
\item $\delta(H) \subset H$
for each homogeneous subspace $H \subset \Lang$
\item $\delta(ab) = \delta(a)b + a\delta(b)$
for all $a,b \in \Lang$
\end{itemize}
Let $\DERIV$ be the set of all graded derivations.
It is a \ff~$\rC$-module\footnote{%
A graded derivation
is determined by its restriction to
$\rC \oplus V \oplus \cc{V}$.
Associated to each basis $v,w \in V$
is a module isomorphism
$\Der(\rC) \oplus \Hom_{\rC}(V,V) \oplus \Hom_{\rC}(\cc{V},\cc{V})
\to \DERIV,
\alpha \oplus \beta' \oplus \beta'' \mapsto \delta$
defined by
$\delta|_{\rC} = \alpha$
and $\delta(v) = \beta'(v), \delta(w) = \beta'(w)$
and $\delta(\cc{v}) = \beta''(\cc{v}), \delta(\cc{w}) = \beta''(\cc{w})$.
}.

\subsection{Graded derivations with shift}\label{sec:grshift}

Let $X \subset \Lang$ be a homogeneous subspace.
By definition,
a map $x: \Lang \to \Lang$ is a graded derivation
with shift $X$, if
for all homogeneous subspaces $H\subset \Lang$:
\begin{itemize}
\item $x$ is linear over $\C$
\item $x(H) \subset XH$
\item $x(ha) = x(h)a + \sigma_{XH}h x(a)$
for all $h\in H$ and $a\in \Lang$.
\end{itemize}
The operator $\sigma_{XH}: HX \to XH$
permutes the leftmost factors $H$ and $X$.

The tensor product of $X$ with $\DERIV$ is denoted
$X\DERIV \subset \LD$. It is canonically isomorphic to\footnote{%
If $\xi \in X$ and $\delta \in \DERIV$,
then $x=\xi \delta \in X\DERIV$
is a graded derivation with shift $X$ because
$x(ha) =
\xi \delta(ha)
=
\xi \delta(h)a + \xi h\delta(a)
=
\xi \delta(h)a + \sigma_{XH} h \xi \delta(a)
= x(h)a + \sigma_{XH} h x(a)$.} the $\rC$-module
of graded derivations with shift $X$.
We identify these modules from now on.


\subsection{Commutators}\label{sec:gencom}

The commutator of graded derivations
$\DB{\,\cdot\,}{\,\cdot\,}: \DERIV\times \DERIV \to\DERIV$
is a $\C$-bilinear map that makes $\DERIV$ into a Lie algebra.

\newcommand{\tbshort}{\tb}
We define a
$\C$-bilinear bracket
$\tb{\,\cdot\,}{\,\cdot\,}: \LD \times \LD \to \LD$.
For all homogeneous subspaces
$X,Y \subset \Lang$
define
$\tb{\,\cdot\,}{\,\cdot\,} : X\DERIV \times Y\DERIV \to XY\DERIV$
by\footnote{%
Equivalently
$\tb{a\delta}{a'\delta'}
= a \delta(a')\delta'
- \delta'(a) a' \delta
+ a a'\DB{\delta}{\delta'}$
for all $a,a'\in \Lang$ and $\delta,\delta'\in \DERIV$.}
\[
\tb{x}{y} \;=\; x\circ y - \sigma_{XY} \circ y \circ x
\]
It makes $\LD$ into a Lie algebra
`with permutations'\footnote{%
This bracket satisfies antisymmetry and Jacobi identities
`with permutations':
\begin{align*}
\tbshort{x}{y} & = -\sigma_{XY \leftarrow YX} \tbshort{y}{x}\\
0 & = \tbshort{x}{\tbshort{y}{z}}
+
\sigma_{XYZ \leftarrow YZX} \tbshort{y}{\tbshort{z}{x}}
+
\sigma_{XYZ \leftarrow ZXY} \tbshort{z}{\tbshort{x}{y}}
\end{align*}
for all $x \in X\DERIV$, $y\in Y\DERIV$, $z \in Z\DERIV$
and all homogeneous subspaces $X,Y,Z \subset \Lang$.
The operators $\sigma_{XYZ \leftarrow YZX}: YZX \to XYZ$ and so forth
are permutation operators.
}.


\subsection{Conjugation}

We define the following
$\rC$-antilinear involutions:
\begin{itemize}
\item $\rC \to \rC,\; f \mapsto\cc{f}$ as in \sref{axioms1}.
\item $\Lang \to \Lang,\; a \mapsto \cc{a}$,
the antilinear algebra automorphism that extends
the antilinear involution 
$f \oplus v \oplus \cc{w} \mapsto \cc{f} \oplus w\oplus \cc{v}$
on $\rC\oplus V\oplus \cc{V}$.
\item
$\DERIV\to \DERIV,\; \delta \mapsto \cc{\delta}$
given by $\cc{\delta}(a) = \cc{\delta(\cc{a})}$.
\item $\LD \to \LD,\; a\delta \mapsto \cc{a \delta}$ given by $\cc{a \delta} = \cc{a}\cc{\delta}$.
\end{itemize}
We refer to all of them as conjugation.

\subsection{Wedge and swap}\label{sec:wedgeswap}

For each homogeneous subspace $H\subset \Lang$
we denote by $H^{\otimes k}=H\cdots H$ its $k$-fold tensor product.
Define $\rC$-linear maps
\begin{align*}
\wedge_k:\; (V\cc{V})^{\otimes k} & \to (V\cc{V})^{\otimes k}\\
\swap_k:\;(\cc{V}V)^{\otimes k}  & \to (V\cc{V})^{\otimes k}
\end{align*}
The first antisymmetrizes the $k$ pairs,
and is normalized by $(\wedge_k)^2 = \wedge_k$.
The second swaps $V$ and $\cc{V}$,
separately in each of the $k$ pairs.
When applied
to tensor products with more factors,
$\wedge_k$ and $\swap_k$
operate on the leftmost factors.

The following $\rC$-antilinear involutions
are particularly useful:
\begin{itemize}
\item $(\VV)^{\otimes k} \to (\VV)^{\otimes k},\; a \mapsto \swap_k \cc{a}$
\item $(\VV)^{\otimes k} \DERIV\to (\VV)^{\otimes k}\DERIV,\; x \mapsto \swap_k \cc{x}$
\end{itemize}
Fixed points of these maps will be referred to as real\footnote{%
Note that $x \in (\VV)^{\otimes k}\DERIV$
is real
if and only if
$x(\cc{a}) = \swap_k \cc{x(a)}$
for all $a \in \Lang$.}.

\subsection{Graded Lie algebra $\D{}$}
\label{sec:gliea}

Let $\D{k} \subset (\VV)^{\otimes k} \DERIV$ be the subset of all $x$
that are both
\begin{itemize}
\item antisymmetric: $x = \wedge_k x$
\item real: $x = \swap_k \cc{x}$
\end{itemize}
Then $\D{k}$ is a \ff~$\rR$-module.
Define 
$\db{\,\cdot\,}{\,\cdot\,}: \D{k} \times \D{\ell}
\to \D{k+\ell}$ by\footnote{%
Equivalently
$\db{a\delta}{a'\delta'}
= (a \wedge \delta(a'))\delta'
- (\delta'(a) \wedge a') \delta
+ (a \wedge a') \DB{\delta}{\delta'}$
for all $\delta,\delta'\in \D{0}$
and all real and antisymmetric
$a \in (\VV)^{\otimes k}$
and $a' \in (\VV)^{\otimes \ell}$,
with
$a \wedge a' = \wedge_{k+\ell}(aa')$ etc.
}
\[
\db{x}{y} \; = \; 
{\wedge_{k+\ell}} \circ {x} \circ {y} -
(-1)^{k\ell} {\wedge_{k+\ell}} \circ {y} \circ {x}
\]
It makes
$\D{} = \bigoplus_{k \geq 0} \D{k}$
into a real graded Lie algebra\footnote{%
Note that $\db{x}{y} = \wedge_{k+\ell} \tb{x}{y}$.
By applying $\wedge$ to the
Jacobi identity `with permutations' for $\tb{\,\cdot\,}{\,\cdot\,}$
we get the graded Jacobi identity for $\db{\,\cdot\,}{\,\cdot\,}$.}.

\section{The graded ideal $\IWeyl{}$}\label{sec:spinideal}
\newcommand{\GEN}[1]{\langle #1 \rangle}

We define submodules $\IWeyl{k} \subset \D{k}$
such that
$\IWeyl{} = \bigoplus_k \IWeyl{k}$ is an ideal,
$\db{\D{}}{\IWeyl{}}\subset \IWeyl{}$.

\subsection{The submodule $\IWeyl{2}$}\label{sec:IWeyl2}
Let $\IWeyl{2} \subset \D{2}$ be the
$\rR$-submodule of all $x$ such that
\begin{align*}
\rule{10mm}{0mm}x(\rC) & = 0\rule{80mm}{0mm}\\
x(V) & \subset 
\;\text{(the submodule of $\VV\VV V$ symmetric in all three $V$)}\\
x(V \wedge V) & = 0
\end{align*}
with $V \wedge V$ the antisymmetric submodule of $VV$.

The conditions
on $x(V)$ and $x(V\wedge V)$ are equivalent
to analogous conditions on
$x(\cc{V})$ and $x(\cc{V}\wedge \cc{V})$,
because elements of $\D{2}$ are real.

\subsection{Creation operators}\label{sec:cr1}

Let $\VVreal\subset \VV$ be the subset of real $c$, that is $c = \swap_1 \cc{c}$.

For every $c \in \VVreal$ and for all $k$ we define the $\rR$-linear map
\begin{align*}
\cre_c\;:\;\D{k} & \to \D{k+1}\\
x & \mapsto \wedge_{k+1}(cx)
\end{align*}
Clearly $\cre_c \circ \cre_c = 0$.
We refer to $\cre_c$ as the creation operator
 of $c$, cf.~\sref{cr2}.

\subsection{The graded submodule $\IWeyl{}$}\label{sec:grsub}

For every submodule $S \subset \D{}$, denote by
$\GEN{S} \subset \D{}$ the smallest submodule
such that $S \subset \GEN{S}$ and such that
$\cre_c(\GEN{S}) \subset \GEN{S}$ for all $c$.
Example: $\D{} = \GEN{\D{0}}$.

We define
\[ \IWeyl{} = \GEN{\IWeyl{2}} \]

Then $\IWeyl{} = \bigoplus_k \IWeyl{k}$
for some\footnote{For $k=2$ this notation is consistent, since
$\GEN{\IWeyl{2}} \cap \D{2} = \IWeyl{2}$.}
 $\IWeyl{k} \subset \D{k}$.
Observe that
\[\IWeyl{0} = \IWeyl{1} = 0\]
We now prove that $\IWeyl{}$ is actually a graded Lie algebra ideal,
$\db{\D{}}{\IWeyl{}} \subset \IWeyl{}$.

\subsection{Lemma}\label{sec:zztzeu1}
If a submodule $S \subset \D{}$ satisfies
$\db{\D{0}}{S} \subset S$, then
$\db{\D{0}}{\GEN{S}} \subset \GEN{S}$.

This follows from:
$\db{\delta}{\cre_c x}
= \cre_c \db{\delta}{x}
+ \cre_{\delta(c)}x$
for all $\delta \in \D{0}$ and $x \in \D{}$.

\subsection{Lemma}\label{sec:zztzeu2}
Let $\FANCY{\ell} \subset \D{\ell}$ be the
set of all $x$ such that
$x(\rC)=0$ and $\wedge_{\ell+1}x(\VV)=0$.
Let $\FANCY{} = \bigoplus_{\ell \geq 0} \FANCY{\ell}$.
Note that $\GEN{\FANCY{}} = \FANCY{}$.

Then
$\db{\GEN{S}}{S'} \subset \GEN{\db{S}{S'}}$
for all submodules $S \subset \D{}$
and $S' \subset \FANCY{}$.

This follows from: $\db{\cre_cx}{x'} = \cre_c \db{x}{x'}$
for all $x \in \D{}$ and $x' \in \FANCY{}$.

\subsection{Lemma}\label{sec:zztzeu3}
If $x \in \IWeyl{2}$
then $x(\rC) = 0$ and $\wedge_3 x(\VV)=0$.

To see this, let $P \subset \VV\VV\VV$
be the submodule symmetric in the three $V$,
and $Q \subset \VV\VV\VV$
the submodule antisymmetric in the three $\VV$.
Elements of $P \cap Q$ are antisymmetric in the three
$\cc{V}$, hence \axV~implies $P \cap Q = 0$.

We have
$\wedge_3 P \subset P$ since
$P$ is spanned by elements of the form
$v\cc{w_1}v\cc{w_2}v\cc{w_3}$.
We also have $\wedge_3 P \subset Q$.
Combining, $\wedge_3 P \subset P\cap Q = 0$.

We must show that $\wedge_3 x(v\cc{w})=0$ for all $x \in \IWeyl{2}$
and $v,w \in V$.
It suffices to show that $\wedge_3 (x(v)\cc{w})=0$;
the case when $x$ hits $\cc{w}$ is analogous.
The definition of $\IWeyl{2}$ in \sref{IWeyl2} implies
$x(v)\cc{w} \in P$,
hence $\wedge_3(x(v)\cc{w})=0$, and we are done\footnote{%
Note that
the condition
$x(V \wedge V) = 0$
in the definition of $\IWeyl{2}$ is not used here.}.

\subsection{The graded submodule $\IWeyl{}$ is a graded ideal}
We have $\db{\D{0}}{\IWeyl{2}} \subset \IWeyl{2}$
directly from \sref{IWeyl2}.
Then \sref{zztzeu1}
implies $\db{\D{0}}{\IWeyl{}} \subset \IWeyl{}$.
We have $\IWeyl{2} \subset \FANCY{}$ by
\sref{zztzeu3}, therefore
$\IWeyl{} = \GEN{\IWeyl{2}} \subset \GEN{\FANCY{}} = \FANCY{}$,
that is,
$\IWeyl{} \subset \FANCY{}$. 
Now \sref{zztzeu2}
gives
$\db{\D{}}{\IWeyl{}}
= \db{\GEN{\D{0}}}{\IWeyl{}} \subset \GEN{\db{\D{0}}{\IWeyl{}}}
\subset \GEN{\IWeyl{}} = \IWeyl{}$.


\subsection{The graded Lie algebra $\Ein{} = \D{}/\IWeyl{}$}
Define the quotient graded Lie algebra
\[\Ein{} = \D{}/\IWeyl{}\]
The induced bracket is denoted
\[
\eb{\,\cdot\,}{\,\cdot\,}\;\;:\;\; \Ein{k}\times \Ein{\ell} \to \Ein{k+\ell}
\]
Explicitly,
$\eb{x \bmod \IWeyl{}}{y \bmod \IWeyl{}} = \db{x}{y} \bmod \IWeyl{}$
for all $x,y\in \D{}$.

Recall from \sref{intro} that we say that $\unk \in \Ein{1}$
solves the Einstein equations if and only if $\eb{\unk}{\unk} = 0$.
In the geometric setting \sref{gs},
each suitably nondegenerate $\unk \in \Ein{1}$
defines a conformal Riemann-Cartan geometry,
see \aref{rc}.
If $\eb{\unk}{\unk}=0$ then this yields
a solution to the standard vacuum Einstein equations.



\section{Formal exponentiation}\label{sec:gaugegroupoid}

In this informal section we discuss
the formal exponentiation of elements of the infinitesimal
gauge group $\Ein{0}$.
These formal exponentials
do actually exist
in the geometric setting \sref{gs},
at least for elements of $\Ein{0}$
with compact support.

Recall that the elements of $\Ein{0}$ are simply
the real graded derivations on $\Lang$.
In fact, since $\IWeyl{0} = 0$, we have
$\Ein{0} = \D{0} \subset \DERIV$.

\subsection{Formal exponentiation of $\delta \in \Ein{0}$}\label{sec:igg}

For every $\delta \in \Ein{0}$ define
\begin{equation*}
e^{\delta}\;:\; \Lang \to \Lang \;,\; 
a_0 \mapsto a_1
\end{equation*}
where $s \mapsto a_s\in \Lang$
is the solution to $\tfrac{\dd}{\dd s} a_s = \delta(a_s)$.

This is an entirely formal definition,
we are just pretending that this differential equation
make sense somehow,
with a unique solution for each $a_0 \in \Lang$.

At least formally:
\begin{itemize}
\item $e^{\delta}$ is linear over $\C$
\item $e^{\delta}(H) \subset H$ for each homogeneous subspace $H\subset \Lang$
\item $e^{\delta}(ab) = (e^{\delta}a)(e^{\delta}b)$ for all $a,b\in \Lang$
\item $e^{\delta}$ commutes with conjugation on $\Lang$
\end{itemize}

In particular, the restriction to $\rR \oplus V \subset \Lang$
is an automorphism that preserves the structure in
\sref{axioms1}, that is,
$e^{\delta}$ is a morphism in the gauge groupoid.
For this reason, we call
$\Ein{0}$ the infinitesimal gauge group. 

\subsection{Formal exponentiation of $\tb{\delta}{\,\cdot\,}$}
For every $\delta \in \Ein{0}$, define
\begin{equation*}
\tbexp{\delta}\;:\; \LD \to \LD \;,\; 
x_0 \mapsto x_1
\end{equation*}
where $s \mapsto x_s \in \LD$
is the solution to $\tfrac{\dd}{\dd s} x_s = \tb{\delta}{x_s}$.

This is, again, an entirely formal definition. It is compatible with \sref{igg},
\begin{equation*}
e^{\delta}(x(a)) = (\tbexp{\delta}x)(e^{\delta}a)
\end{equation*}
i.e.~if
$\tfrac{\dd}{\dd s}x_s = \tb{\delta}{x_s}$
and
$\tfrac{\dd}{\dd s}a_s = \delta(a_s)$,
then
$b_s = x_s(a_s)$ solves $\tfrac{\dd}{\dd s}b_s = \delta(b_s)$.

Formally, the map $\tbexp{\delta}$
induces a map $\D{}\to \D{}$ that leaves the bracket 
$\db{\,\cdot\,}{\,\cdot\,}$ invariant,
and a map $\Ein{} \to \Ein{}$
that leaves the bracket $\eb{\,\cdot\,}{\,\cdot\,}$ invariant.
In particular, if $\unk\in \Ein{1}$ solves the Einstein equations
then so does $\tbexp{\delta}\unk$.

\addtocontents{toc}{~\emph{Gauges algorithm}\par}

\section{Algorithm specification}\label{sec:algspec}

We consider an algorithm with
\begin{center}
\begin{tabular}{r@{\quad}l}
{\bf input:} & a $\rC$-Hermitian map, see \sref{alginp}\\
{\bf output:} & a sequence of $\rR$-bilinear maps, see \sref{algout}
\end{tabular}
\end{center}
Thus the algorithm performs an abstract task,
but its output is precisely what we need in
\sref{lingr} and \sref{nonlingr}
to gauge-fix and to get wave equations (symmetric hyperbolic equations).
The algorithm is concisely specified in \sref{alginp} and \sref{algout},
which suffice to read \sref{lingr} and \sref{nonlingr}.
The algorithm itself is in \sref{algimpl}.

The algorithm
is equivariant under the action of the gauge groupoid.
That is, the algorithm commutes with the
canonical action of the gauge groupoid on the
input and output of the algorithm, cf.~remarks in \sref{volform}.

\subsection{Input of the algorithm}\label{sec:alginp}
The 
input is a map
\[
\gauge\;\;:\;\;
\cc{V}\DERIV\;\times\; \cc{V}\DERIV\;\to\;\rC
\]
such that:
\begin{itemize}
\item[\ixherm] $\gauge$ is $\rC$-Hermitian,
namely $\gauge$ is
$\rC$-antilinear in the first argument,
$\rC$-linear in the second argument,
and satisfies $\cc{\gauge(x,y)} = \gauge(y,x)$.
\item[\ixker]
$\gauge(\,\cdot\,,\KERN) = 0$, where\footnote{%
The `kernel' $\KERN$ is
closely related to the ideal
$\IWeyl{}$, see \sref{auxr}.
} $\KERN \subset \cc{V}\DERIV$
is the $\rC$-submodule of elements $x$ with
\begin{align*}
x(\rC) & = 0\\
x(V) & = 0\\
x(\cc{V}) & \subset
\;\text{(the symmetric submodule of $\cc{V}\cc{V}$)}\\
x(\cc{V} \wedge \cc{V}) & = 0
\end{align*}
\item[\ixpos]
$\gauge: (\cc{V}\DERIV/\KERN)
\times (\cc{V}\DERIV/\KERN) \to \rC$ is positive definite.
\end{itemize}

We say that a Hermitian form on a \ff~$\rC$-module
is positive definite
if there is an orthonormal basis\footnote{That is,
there is a module isomorphism to $\rC^n$
for some $n$
that makes the
Hermitian form become the standard form
$\rC^n\times \rC^n \to \rC, ((g_1,\ldots,g_n),(f_1,\ldots,f_n))
\mapsto \cc{g_1}f_1 + \ldots + \cc{g_n}f_n$.
}; same definition for a symmetric bilinear form on an \ff~$\rR$-module.
To make things go smoothly,
we adopt the axiom:
\begin{itemize}
\item[\axpos] Given any positive definite $\rC$-Hermitian form
on a \ff~$\rC$-module $M$,
and a \ff~submodule $M' \subset M$ with \ff~complement\footnote{%
That is, $M = M' \oplus M''$ where $M''$ is also \ff.} in $M$,
then the restriction of the
Hermitian form to $M'$ is also positive definite.
Same for positive definite symmetric forms on a \ff~$\rR$-module.
\end{itemize}
It holds in the geometric setting \sref{gs},
like the axioms in \sref{axioms1}.

\subsection{Output of the algorithm}\label{sec:algout}
Set $\EinG{-1} = 0$.

The algorithm returns sequentially for $k=-1,0,1,2,\ldots$
an $\rR$-bilinear map
\[
\bil{k} \;\;:\;\; \EinG{k} \;\times\; \Ein{k+1} \;\to\; \rR
\]
such that for all
\[c \;\;\in\;\; \VVpos = \{v\cc{v}+w\cc{w}\mid \textmd{$v,w$ a basis for $V$}\}\]
we have\footnote{%
The operator $\cre_c: \D{k}\to \D{k+1}$ is defined in \sref{cr1}
or \sref{cr2}.
It satisfies $\cre_c(\IWeyl{k}) \subset \IWeyl{k+1}$.
}:
\begin{itemize}
\item[\kk{\oxsp}{k}] $\EinG{k} \times \EinG{k} \to \rR,
\;(x,y) \mapsto \bil{k}(x,\cre_cy)$ is symmetric and positive definite\footnote{The submodule
$\EinG{k}\subset \Ein{k}$ is already known
from \kk{\oxds}{k-1},
and is \ff~by \kk{\oxff}{k-1}.
}.
\item[\kk{\oxds}{k}]
Define
\[\EinG{k+1} = \{x \in \Ein{k+1}\mid \bil{k}(y,x)=0~\textmd{for all $y \in \EinG{k}$}
\}\footnote{%
Definitions such as this one will from now on be abbreviated by
just $\bil{k}(\EinG{k},\EinG{k+1}) = 0$.}\] Then
 \[
\Ein{k+1} = \cre_c(\EinG{k}) \oplus \EinG{k+1}
\]
is an internal direct sum decomposition,
and $\cre_c: \EinG{k} \to \Ein{k+1}$ is injective.
\item[\kk{\oxff}{k}] $\EinG{k+1}$ is \ff.
\end{itemize}

This information is all that one needs to read \sref{lingr}
and \sref{nonlingr},
where the bilinear forms $\bil{k}$
are used to construct symmetric hyperbolic operators.

\subsection{The `gauge condition'}\label{sec:tgc}
The algorithm in \sref{algimpl} yields in particular $\EinG{0} = \Ein{0}$ and
\[\bil{0}(\delta,\unk) \;=\; \RE\,\gauge(\cc{\unk_1}\delta,\unk_2\unk_3)
\qquad \textmd{for all $\delta \in \Ein{0}$, $\unk \in \Ein{1}$}\]
Here $\unk = \sum_i \unk_{1i}\unk_{2i}\unk_{3i}$
is any decomposition
with $\unk_{1i} \in V$, $\unk_{2i}\in \cc{V}$, $\unk_{3i}\in \DERIV$,
and we suppress the sum over $i$ on the right hand side, cf.~\sref{uncon}.

This determines $\EinG{1}$ through $\bil{0}(\Ein{0},\EinG{1}) = 0$.
Informally:
$\EinG{1}$ is determined by as many linear conditions
as there are degrees of freedom in the infinitesimal gauge group $\Ein{0}$.
Therefore $\unk \in \EinG{1}$
makes naive sense as a `gauge condition'.
The algorithm shows that it makes actual sense. 

\subsection{Remarks}

\emph{\kk{\oxsp}{k} implies \kk{\oxds}{k}.}
Fix any $c \in \VVpos$. We must show that for every
$x \in \Ein{k+1}$ there is a unique $y \in \EinG{k}$
such that $x-\cre_cy \in \EinG{k+1}$,
that is such that $\bil{k}(\,\cdot\,,x-\cre_cy) = 0
\in \Hom_{\rR}(\EinG{k},\rR)$.
There is unique such $y$ by \kk{\oxsp}{k}.

\emph{\kk{\oxds}{k-1} and \kk{\oxds}{k} imply \kk{\oxff}{k}.}
Fix any $c \in \VVpos$.
The quotient
$\Ein{k+1}/\cre_c(\Ein{k})$ is \ff:
this space does not involve $\gauge$,
and a basis can be exhibited explicitly.
By \kk{\oxds}{k-1} we have
$\cre_c(\Ein{k}) = \cre_c(\EinG{k})$.
By \kk{\oxds}{k} we have
$\EinG{k+1} \cong \Ein{k+1}/\cre_c(\EinG{k})$.
Combining, $\EinG{k+1} \cong \Ein{k+1}/\cre_c(\Ein{k})$,
and \kk{\oxff}{k} follows.

It follows from \oxds~that
\begin{alignat*}{10}
0 &\xrightarrow{\;\;\cre_c\;\;}\;& \Ein{0}
&\xrightarrow{\;\;\cre_c\;\;}\;& \Ein{1} 
&\xrightarrow{\;\;\cre_c\;\;}\;& \Ein{2}
&\xrightarrow{\;\;\cre_c\;\;}\;& \Ein{3}
&\xrightarrow{\;\;\cre_c\;\;}\;& \Ein{4}
&\xrightarrow{\;\;\cre_c\;\;}\;& 0
\end{alignat*}
is an exact sequence for all $c \in \VVpos$.

The graded Lie algebra bracket
$\eb{\,\cdot\,}{\,\cdot\,}$ does not appear in the algorithm.
Its stand-in is the creation operator $\cre_c$.
The two are linked by the identity
\begin{equation}\label{eq:comi}
\eb{\unk}{f\,\cdot\,} - f\eb{\unk}{\,\cdot\,} = \cre_{\unk(f)}
\;\;\;:\;\;\; \Ein{} \to \Ein{}
\end{equation}
that holds for all $\unk \in \Ein{1}$ and $f \in \rR$.


\section{Preliminaries}

\subsection{Auxiliary volume form $\eps$}\label{sec:volform}

A volume form on $V$ is a map
$\eps : V\times V \to \rC$
that is
$\rC$-bilinear and antisymmetric,
and such that
$\eps(v,w)$ has a multiplicative inverse in $\rC$ for every basis $v,w \in V$.
By \axV~such a volume form  $\eps$ exists and is unique up to
scaling
by an invertible element of $\rC$.
We define $\cc{\eps}: \cc{V}\times \cc{V}\to \rC$
by $\cc{\eps}(\cc{v},\cc{w}) = \cc{\eps(v,w)}$.

From now on we use an auxiliary volume form $\eps$.
We emphasize that the choice of $\eps$
does not affect the output \sref{algout} of the algorithm:
\begin{itemize}
\item The spaces $\EinG{k}$ do not depend on $\eps$.
\item The maps $\bil{k}$ scale homogeneously:
$\bil{k}/(\eps\cc{\eps})^k$ does not depend on $\eps$.
\end{itemize}
All objects in the algorithm
scale homogeneously\footnote{%
Look-up list for the scaling of
some of the main objects used in
\sref{auxr}, \sref{auxbil}, \sref{algimpl}:
\begin{align*}
\cre_c & \propto 1 &
\rac{2k} & \propto \eps^k\cc{\eps}^{k-1} &
\je & \propto (\eps\cc{\eps})^{-1} &
\BG{k} & \propto (\eps \cc{\eps})^k &
\auxx{k} & \propto \eps\cc{\eps}\\
\ann_c & \propto \eps \cc{\eps}&
\rac{2k+1} & \propto \eps^k\cc{\eps}^k &
\km & \propto \eps\cc{\eps} &
\bil{k} & \propto (\eps\cc{\eps})^k
\end{align*}
}.
Therefore $\eps$ is not additional structure,
and the algorithm is equivariant under the original gauge groupoid.

\subsection{Unconventional notation}\label{sec:uncon}
In this notation,
an explicit definition of $\wedge_2: (\VV)^{\otimes 2}\to(\VV)^{\otimes 2}$
in \sref{wedgeswap} is
$\wedge_2a = \tfrac{1}{2}(a_1a_2a_3a_4-a_3a_4a_1a_2)$.
The understanding is that $a \in (\VV)^{\otimes 2}$ is a sum of products,
$a = \sum_i a_{1i}a_{2i}a_{3i}a_{4i}$,
and that the index $i$ and the sum are
conveniently suppressed on the right hand side.

In general,
suppose $X = X_1 \otimes_{\rC} \cdots \otimes_{\rC} X_n$
and suppose $f: X_1 \times \cdots \times X_n \to Y$ is $\rC$-linear
in each argument.
Then for each $x\in X$ the notation
$f(x_1,\ldots,x_n)$
is an abbreviation for the finite sum $\sum_i f(x_{1i},\ldots,x_{ni})$,
with
$x = \sum_i x_{1i}\cdots x_{ni}$
and $x_{ji} \in X_j$.
The result does not depend on how $x$ is decomposed.

\subsection{Creation and annihilation operators}\label{sec:cr2}

For every $c \in \VVreal$  define $\rR$-linear maps
\begin{align*}
\cre_c\;:\;\D{k} & \to \D{k+1}\\
x & \mapsto \wedge_{k+1}(cx)\displaybreak[0]\\
\rule{0pt}{14pt}\ann_c\;:\;\D{k} & \to \D{k-1} \\
x & \mapsto \tfrac{1}{2}\,k\,\eps(c_1,x_1) \cc{\eps}(c_2,x_2)x_3\ldots x_{2k}x_{2k+1}
\qquad \textmd{(using \sref{uncon})}
\end{align*}
We refer to $\cre_c$ and $\ann_c$ as the creation and annihilation operators
of $c$.
If $\cre,\ann$ and $\cre',\ann'$ are the creation and annihilation operators
of $c$ and $c'$, then
\begin{alignat*}{6}
\cre& \cre' &&\;+\;& \cre'& \cre && \;=\; 0\\
\ann& \ann' &&\;+\;& \ann'& \ann && \;=\; 0\\
\ann& \cre' &&\;+\;& \cre'& \ann && \;=\; \tfrac{1}{2}\,\eps(c_1,c_1')\cc{\eps}(c_2,c_2')\,\mathbbm{1}
\qquad \textmd{(using \sref{uncon})}
\end{alignat*}
where
$\mathbbm{1}: \D{} \to \D{}$
is the identity.

The creation operators $\cre_c$ descend to $\Ein{} \to \Ein{}$ by \sref{grsub},
but not so the $\ann_c$.

\section{Auxiliary linear maps}\label{sec:auxr}
This section uses the space $\KERN$ in \ixker.
It does not use $\gauge$.
\subsection{The maps $\rac{k}$}\label{sec:racmaps}

We define $\rC$-linear maps
\[
\rac{k}\;:\; (\VV)^{\otimes k}\DERIV \to \begin{cases} \cc{V}\cc{V}\DERIV & \textmd{if $k$ is even}\\
\VV\DERIV & \textmd{if $k$ is odd}
\end{cases}
\]
Set $\rac{k}=0$ when $k<0$ and set
\[
\rac{0}(x) = \tfrac{1}{\cc{\eps}(\cc{v},\cc{w})}
(\cc{w}\cc{v}-\cc{v}\cc{w})x
\]
where $v,w \in V$ is any basis; the definition does not depend
on the choice of the basis. For $k>0$ set recursively
\[
\rac{k}(x)
=
\begin{cases}
\eps(x_1,X_1)x_2X_2X_3 & \textmd{if $k$ is even}\\
\cc{\eps}(x_2,X_1)x_1X_2X_3 & \textmd{if $k$ is odd}\\
\end{cases}
\]
where $X = X_1X_2X_3 = \rac{k-1}(x_3x_4\cdots x_{2k}x_{2k+1})$,
using the notation in \sref{uncon}.

Explicitly\footnote{%
Example:
$\rac{1}(x) = \cc{\eps}(x_2,X_1)x_1X_2X_3$
where $X_1X_2X_3 = \rac{0}(x_3) = \smash{\frac{1}{\cc{\eps}(\cc{v},\cc{w})}}
(\cc{w}\cc{v}-\cc{v}\cc{w})x_3$, hence
$\rac{1}(x) = \smash{\frac{1}{\cc{\eps}(\cc{v},\cc{w})}}
(\cc{\eps}(x_2,\cc{w})x_1\cc{v}x_3 - \cc{\eps}(x_2,\cc{v})x_1\cc{w}x_3)
= x_1x_2x_3$.
For the last step, note that
$\eps(u,v)w + \eps(v,w)u + \eps(w,u)v = 0$
for all $u,v,w \in V$ by \axV.
}
\begin{alignat*}{6}
\rac{1}(x) & = x_1x_2x_3 = x\\
\rac{2}(x) & = \eps(x_1,x_3)x_2x_4x_5\\
\rac{3}(x) & = \cc{\eps}(x_2,x_4)\eps(x_3,x_5)x_1x_6x_7\\
\rac{4}(x) & = \eps(x_1,x_3) \cc{\eps}(x_4,x_6)\eps(x_5,x_7)x_2x_8x_9
\end{alignat*}
and so forth.

\subsection{The element $\je$}
Let $v,w\in V$ be any basis and set
\[
\je = \tfrac{1}{2\,\eps(v,w)\cc{\eps}(\cc{v},\cc{w})}(v\cc{v}w\cc{w}-w\cc{v}v\cc{w}-v\cc{w}w\cc{v}+w\cc{w}v\cc{v})
\;\in\;\VV\VV
\]
The element $\je$ does not depend on the choice of the basis.
It is antisymmetric in the two $V$,
antisymmetric in the two $\cc{V}$, and real, $\je = \swap_2 \cc{\je}$.
The direct sum
$\je \D{k} \oplus \D{k+2}$
is well-defined as an internal direct sum in $(\VV)^{\otimes(k+2)}\DERIV$.

\subsection{Key lemma}\label{sec:klr}
For all $c \in \VVreal$ we have
\begin{align*}
\rac{k}(x) & = - \rac{k+2}(\je x) \\
\rac{k}(cx) & = - \rac{k+2}(c\je x)\\
\rac{k+1}(\cre_cy) - \rac{k-1}(\ann_cy) & = \rac{k+1}(cy)
\qquad \textmd{if $y \in \D{k}$}
\end{align*}
One can prove these by expanding everything in a basis for $V$.

\subsection{The map $\km$}

Define an $\rR$-linear map
$\km: \{x \in \D{3}\mid x(\rC)=0\} \to \D{1}$ by
\begin{align*}
\km(x)(\rC) &=0\\
\km(x)(v) &= \rac{3}(x)(v)\\
\km(x)(\cc{v}) &= -\rac{3}(x)(\cc{v})
\end{align*} 
for all $v \in V$.
By direct verification,
this yields an element of $\D{1}$. 

\subsection{Lemma}\label{sec:binsr}
We have 
$\rac{2}(\IWeyl{2}) \subset \cc{V} \KERN$
and\footnote{%
Let $x \in \D{3}$ and $x(\rC)=0$
and $y = \rac{3}((\je \km+\mathbbm{1})(x))$.
Then
$y = -\km(x) + \rac{3}(x)$ using \sref{klr}, and
$y(\rC)=0$, $y(v) = 0$, $y(\cc{v}) = 2\rac{3}(x)(\cc{v})$
for all $v \in V$. If $x \in \IWeyl{3}$
then $y(\cc{v}) \in V\cc{V}\cc{V}$ is symmetric in the two $\cc{V}$,
and $y(\cc{V}\wedge \cc{V})=0$.
}
$\rac{3}((\je\km+\mathbbm{1})(\IWeyl{3})) \subset V\KERN$.

\subsection{Lemma}\label{sec:INJ}
Introduce the \ff~$\rR$-modules
\begin{align*}
\XNN{0} & = \cc{V}\cc{V} \DERIV/\cc{V}\KERN &
\XNN{1} & = \VV\DERIV/V\KERN
\intertext{as well as}
\msp{0} & = \D{0}&
\msp{2} & = (\je \D{0} \oplus \D{2})/\IWeyl{2}\\
\msp{1} & = \D{1}&
\msp{3} & = (\je \D{1} \oplus \D{3})/((\je \km+\mathbbm{1})(\IWeyl{3}))
\end{align*}

Then for $k=0,1,2,3$ the map
$\rac{k}$ descends to a map $\msp{k} \to \XNN{k \bmod 2}$
that is injective\footnote{%
Recall the definitions of $\KERN$ in \sref{alginp},
and of $\rac{k}$ in \sref{racmaps}.
Injectivity for $k=0$:
Suppose $x \in \D{0}$ and $\rac{0}(x) \in \cc{V}\KERN$,
then $x(\rC)=x(V)=0$, hence $x=0$ since $x$ is real.
Injectivity for $k=1$ is similar.
Injectivity for $k=2$:
Suppose $x \in \D{0}$
and $y \in \D{2}$
and $\rac{2}(\je x + y) \in \cc{V}\KERN$,
then $x(\rC)=x(V)=0$,
hence $x=0$ since $x$ is real,
hence $\rac{2}(y) \in \cc{V}\KERN$,
hence $y(\rC)=y(V \wedge V)=0$
by reality of $y$,
and $y(V) \subset \VV\VV V$ is
(in the submodule) symmetric in the first two $V$,
hence reality of $y$ implies that $y(\cc{V}) \subset \VV\VV\cc{V}$
is symmetric in the first two $\cc{V}$,
hence antisymmetric in the two $V$ by $\wedge_2y=y$,
hence also symmetric in the last two $\cc{V}$
by $\rac{2}(y) \in \cc{V}\KERN$,
hence symmetric in all three $\cc{V}$,
and $y \in \IWeyl{2}$ as required.
Injectivity for $k=3$:
Suppose $x \in \D{1}$ and $y \in \D{3}$
and $\rac{3}(\je x + y) \in V \KERN$,
then
$-x + \rac{3}(y) \in V \KERN$
with $x \in \D{1}$
and $\rac{3}(y)\in i\D{1}$,
then
$x(\rC)=y(\rC)=x(V\wedge V)=y(V\wedge V)=0$,
and $x(v) = \rac{3}(y)(v)$ for all $v \in V$,
hence $x(\cc{v}) = -\rac{3}(y)(\cc{v})$ by reality of $x$ and $y$,
hence $x = \km(y)$
and $\rac{3}(y)(\cc{V}) \subset V\cc{V}\cc{V}$
is symmetric in the two $\cc{V}$
and $\rac{3}(y)(\cc{V} \wedge \cc{V})=0$,
hence (without proof) $y \in \IWeyl{3}$.
},
and without proof
its image has \ff~complement in $\XNN{k \bmod 2}$.

\section{Auxiliary bilinear maps}\label{sec:auxbil}

The algorithm in \sref{algimpl}
is based on the auxiliary bilinear maps
$\BG{k}$ introduced in this section.
They are constructed using $\gauge$.

\newcommand{\baux}{\texttt{B}_{\ast}}

\subsection{The map $\baux$}\label{sec:baux}
Define the $\rR$-bilinear map
$\baux: \cc{V}\cc{V}\DERIV\times \VV\DERIV\to\rR$ by,
using \sref{uncon},
\[
\baux(x,y) \;=\;\RE 
\,\eps(\cc{x_1},y_1)\,\gauge(x_2x_3,y_2y_3)
\]
Note that $\baux(\cc{V}\KERN,\,\cdot\,)=0$ and $\baux(\,\cdot\,,V\KERN)=0$
by \ixker.

\subsection{The bilinear maps $\BG{k}$}
Define $\rR$-bilinear maps
$\BG{k}: (\VV)^{\otimes k}\DERIV \times (\VV)^{\otimes(k+1)}\DERIV \to \rR$ by
\[
\BG{k}(x,y) \;=\;
\begin{cases}
-\baux(\rac{k}(x),\rac{k+1}(y)) & \textmd{if $k$ is even}\\
+\baux(\rac{k+1}(y),\rac{k}(x)) & \textmd{if $k$ is odd}
\end{cases}
\]

\subsection{Lemma} \label{sec:klo}
By \sref{klr}
we have for all $c \in \VVreal$:
\begin{align*}
\BG{k}(x,y) &= \BG{k+1}(y,\je x)\\
\BG{k}(cx,y) & = \BG{k+1}(y,c\je x)\\
\BG{k}(x,\cre_cy) 
+
\BG{k-1}(\ann_cy,x)
& =
\BG{k}(x,cy)
\qquad \textmd{if $y \in \D{k}$}
\end{align*}

\subsection{Lemma}\label{sec:binsb}

By \sref{binsr} the following all vanish:
\begin{align*}
&&
\BG{0}(\,\cdot\,,\IWeyl{1}) &  &
\BG{1}(\,\cdot\,,\IWeyl{2}) & &
\BG{2}(\,\cdot\,,(\je \km+\mathbbm{1})(\IWeyl{3})) & \\
\BG{0}(\IWeyl{0},\,\cdot\,) &  &
\BG{1}(\IWeyl{1},\,\cdot\,) & &
\BG{2}(\IWeyl{2},\,\cdot\,) &  &
\BG{3}((\je \km+\mathbbm{1})(\IWeyl{3}),\,\cdot\,) & 
\end{align*}

\newcommand{\haux}[1]{\texttt{H}_{#1}}
\newcommand{\chaux}[1]{\cc{\texttt{H}}_{#1}}
\subsection{The map $\haux{c}$}\label{sec:haux}
For every $c \in \VVreal$ define
$\haux{c}:V\times V \to \rC$ and
$\chaux{c}:\cc{V}\times \cc{V}\to \rC$
by
\begin{align*}
\haux{c}(x,y) &= \cc{\eps}(c_2,\cc{x})\eps(c_1,y)
&
\chaux{c}(x,y) &= \eps(c_1,\cc{x})\cc{\eps}(c_2,y)
\end{align*}
They are $\rC$-Hermitian, and we have
$\chaux{c}(\cc{x},\cc{y}) = \cc{\haux{c}(x,y)}$.

If $c \in \VVpos$ then they are positive definite:
If $c= v\cc{v}+w\cc{w}$ for a basis $v,w\in V$,
then an $\haux{c}$-orthonormal basis
is given by $v/\eps(w,v),w/\eps(v,w)$.

By multiplying Hermitian forms, we get
$\chaux{c}\cdot \gauge: \cc{V}\cc{V}\DERIV \times \cc{V}\cc{V}\DERIV \to \rC$
and
$\haux{c}\cdot \gauge: \VV\DERIV \times \VV\DERIV \to \rC$.
If $c \in \VVpos$,
then they
induce positive definite $\rC$-Hermitian forms on the spaces
$\XNN{0}$ and $\XNN{1}$ in \sref{INJ}, by \ixker~and \ixpos.

\subsection{Lemma}\label{sec:klp}
For all $c \in \VVreal$ and all $x,y\in (\VV)^{\otimes k}\DERIV$ we have
\[
\BG{k}(x,cy) \;=\;
\begin{cases}
\RE\;(\chaux{c}\cdot \gauge)(\rac{k}(x),\rac{k}(y)) & \textmd{if $k$ is even}\\
\RE\;(\haux{c}\cdot \gauge)(\rac{k}(x),\rac{k}(y)) & \textmd{if $k$ is odd}\\
\end{cases}
\]
In particular, $\BG{k}(x,cy) = \BG{k}(y,cx)$.


\subsection{Lemma}\label{sec:POS}
For $k=0,1,2,3$ and for all $c \in \VVpos$ the $\rR$-bilinear map
\[
\BG{k}(\,\cdot\,,c\,\cdot\,)\;:\; \msp{k} \times \msp{k} \to \rR
\quad \text{is symmetric and positive definite}
\]
by \sref{INJ} and \sref{haux} and \sref{klp} and \axpos.


\newcommand{\LT}[1]{\rule{30mm}{0pt}\mathllap{#1}}
\newcommand{\RT}[1]{\mathrlap{#1}\rule{70mm}{0pt}}
\newcommand{\CT}[1]{\mathrlap{\textmd{#1}}\rule{30mm}{0pt}}

\section{The algorithm}\label{sec:algimpl}

Recall that $\Ein{k} = \D{k}/\IWeyl{k}$.
Throughout this algorithm,
$\DG{k} \subset \D{k}$ is the preimage of
$\EinG{k} \subset \Ein{k}$ under $\D{k} \to \Ein{k}$.
Therefore $\IWeyl{k} \subset \DG{k}$
and $\EinG{k} = \DG{k}/\IWeyl{k}$.

We sequentially construct maps $\bil{k}$
as in \sref{algout}.

\subsection{The map $\bil{-1}$}
Since $\EinG{-1}=0$ we must set $\bil{-1} = 0$.
Then \kk{\oxsp}{-1} holds.

The definition of $\EinG{0}$
in \kk{\oxds}{-1} yields $\EinG{0} = \Ein{0}$, equivalently
\begin{equation}\label{ed0}
\DG{0} = \D{0}
\end{equation}

\subsection{The map $\bil{0}$}

Set $\bil{0} = \BG{0}$ on $\DG{0} \times \D{1}$.
It vanishes when we insert $\IWeyl{0}$ in the first slot,
or $\IWeyl{1}$ in the second slot, by \sref{binsb}.
 Hence
$\bil{0}$ descends to $\EinG{0} \times \Ein{1}$,
as required.

If $x,y \in \DG{0}$ then
\begin{alignat*}{6}
\LT{\bil{0}(x,\cre_cy)}
& = \RT{\BG{0}(x,\cre_cy)}\\
& = \RT{\BG{0}(x,cy) - \BG{-1}(\ann_c y,x)} && \CT{by \sref{klo}}\\
& = \RT{\BG{0}(x,cy)} && \CT{by $\BG{-1} = 0$}
\end{alignat*}
Now \kk{\oxsp}{0}
follows from \sref{POS} and $\EinG{0} = \Ein{0} = \msp{0}$.

The definition of $\EinG{1}$ in \kk{\oxds}{0}
is equivalent to, using \eqref{ed0},
\begin{equation}\label{ed1}
\BG{0}(\D{0},\DG{1}) = 0
\end{equation}


\subsection{The map $\bil{1}$}
Set $\bil{1} = \BG{1}$ on $\DG{1} \times \D{2}$.
It vanishes when we insert $\IWeyl{1}$ in the first slot,
or $\IWeyl{2}$ in the second slot, by \sref{binsb}. Hence
$\bil{1}$ descends to $\EinG{1} \times \Ein{2}$,
as required.

If $x,y \in \DG{1}$ then
\begin{alignat*}{6}
\LT{\bil{1}(x,\cre_cy)}
& = \RT{\BG{1}(x,\cre_cy)}\\
& = \RT{\BG{1}(x,cy) - \BG{0}(\ann_c y,x)} && \CT{by \sref{klo}}\\
& = \RT{\BG{1}(x,cy)} && \CT{by \eqref{ed1}}
\end{alignat*}
Now \kk{\oxsp}{1} follows from
\sref{POS}
and axiom \axpos, because
$\EinG{1} \subset \Ein{1} = \msp{1}$
has \ff~complement in $\msp{1}$ by \kk{\oxds}{0}.

The definition of $\EinG{2}$ in \kk{\oxds}{1}
is equivalent to
\begin{equation}\label{ed2}
\BG{1}(\DG{1},\DG{2}) = 0
\end{equation}


\subsection{The auxiliary map $\auxx{2}$}

Defining $\bil{2}$ in a way analogous to $\bil{0}$
and $\bil{1}$ fails. We need an extra step.

We show that there is a unique
$\rR$-linear map $\auxx{2}: \DG{2} \to \D{0}$ such that
\begin{equation}\label{eq:a2}
\BG{1}(\D{1},(\je \auxx{2}+\mathbbm{1})(\DG{2}))=0
\end{equation}
Equivalently:
$\BG{0}(\auxx{2}(x),\,\cdot\,) + \BG{1}(\,\cdot\,,x)
= 0 \in \Hom_{\rR}(\D{1},\rR)$,
for all $x \in \DG{2}$.

Let us first discuss the idea. Consider
$\BG{0}$ and $\BG{1}$ as maps
\begin{alignat*}{6}
\D{0} &\; \times\; && \D{1} && \to \rR\\
\D{1} &\; \times\; && \DG{2} && \to \rR
\end{alignat*}
By \eqref{ed1} and \eqref{ed2} they descend to
\begin{alignat*}{6}
\D{0} & \;\times\; && (\D{1}/\DG{1})\; && \to \rR\\
(\D{1}/\DG{1}) & \;\times\; && \DG{2} && \to \rR
\end{alignat*}
Suppose the first pairing is nondegenerate.
We would get a map
from $\DG{2}$ to $\Hom_{\rR}(\D{1}/\DG{1},\rR)$,
and then to $\D{0}$. This map would be $-\auxx{2}$.

Here is the complete argument.
Let $x \in \DG{2}$. We have to show that there is a unique
$y \in \D{0}$ such that
$\BG{0}(y,\,\cdot\,) + \BG{1}(\,\cdot\,,x) = 0 \in \Hom_{\rR}(\D{1},\rR)$.
The condition holds automatically on $\DG{1}$
by \eqref{ed1}, \eqref{ed2}.
Fix $c \in \VVpos$
and use $\D{1} = \cre_c(\DG{0}) \oplus \DG{1}$,
which follows from \kk{\oxds}{0}, to reduce the problem
to
$\BG{0}(y,\cre_c\,\cdot\,) + \BG{1}(\cre_c\,\cdot\,,x)
= 0 \in \Hom_{\rR}(\D{0},\rR)$. There is a unique such $y \in \D{0}$ by
\kk{\oxsp}{0}.


\subsection{The map $\bil{2}$}

Set $\Auxx{2} = \je \auxx{2} + \mathbbm{1}: \DG{2} \to (\VV)^{\otimes 2}\DERIV$.
Note that $\Auxx{2} = \mathbbm{1}$ on $\IWeyl{2}$,
by \sref{binsb}.

Set $\bil{2} = \BG{2}(\Auxx{2}(\,\cdot\,),\,\cdot\,)$
on $\DG{2} \times \D{3}$.
It vanishes when we insert $\IWeyl{2}$ in the first slot
by \sref{binsb}.
To see that
$\bil{2}$ descends to $\EinG{2} \times \Ein{3}$,
we must show
$\bil{2}(\,\cdot\,,\IWeyl{3})=0$.

Let $x \in \DG{2}$, let $c \in \VVreal$, and let $y \in \D{2}$.
Then
\begin{alignat*}{6}
\LT{\bil{2}(x,\cre_cy)} &
= \RT{\BG{2}(\Auxx{2}(x),\cre_cy)}\\
& = \RT{\BG{2}(\Auxx{2}(x),cy) - \BG{1}(\ann_cy,\Auxx{2}(x)0)}
&& \CT{by \sref{klo}}\\
& = \RT{\BG{2}(\Auxx{2}(x),cy)} && \CT{by \eqref{eq:a2}}
\end{alignat*}
This calculation is used twice:
\begin{itemize}
\item We continue with $y \in \IWeyl{2}$:
\begin{alignat*}{6}
\LT{\bil{2}(x,\cre_cy)}
& =
\RT{\BG{2}(\Auxx{2}(x),cy)}\\
& = \RT{\BG{2}(y,c\Auxx{2}(x))} && \CT{by \sref{klp}}\\
& = \RT{0} && \CT{by \sref{binsb}}
\end{alignat*}
Since $\IWeyl{3}$
is spanned by such $\cre_cy$, we 
get $\bil{2}(\,\cdot\,,\IWeyl{3})=0$, as required.
\item We continue with $y \in \DG{2}$:
\begin{alignat*}{6}
\LT{\bil{2}(x,\cre_cy)} & = \RT{\BG{2}(\Auxx{2}(x),cy)}\\
& = \RT{\BG{2}(\Auxx{2}(x),c\Auxx{2}(y)) - \BG{2}(\Auxx{2}(x),c\je \auxx{2}(y))}\\
& = \RT{\BG{2}(\Auxx{2}(x),c\Auxx{2}(y)) - \BG{1}(c\auxx{2}(y),\Auxx{2}(x))} && \CT{by \sref{klo}}\\
& = \RT{\BG{2}(\Auxx{2}(x),c\Auxx{2}(y))} && \CT{by \eqref{eq:a2}}
\end{alignat*}
and we have used
$c \auxx{2}(y)\in \D{0}$.
Now \kk{\oxsp}{2} follows from
\sref{POS}
and axiom \axpos, because
$\Auxx{2}$ 
descends to an injective map
$\EinG{2} \to \je \D{0} \oplus \Ein{2} = \msp{2}$
with \ff~complement by \kk{\oxsp}{1}.
\end{itemize}

The definition of $\EinG{3}$ in \kk{\oxds}{2}
is equivalent to
\begin{equation}\label{ed3}
\BG{2}(\Auxx{2}(\DG{2}),\DG{3}) = 0
\end{equation}


\subsection{The auxiliary map $\auxx{3}$}

There is a unique
$\rR$-linear map $\auxx{3}: \DG{3} \to \D{1}$ such that
\begin{equation}\label{eq:a3}
\BG{2}(\je \D{0} \oplus \D{2},(\je \auxx{3}+\mathbbm{1})(\DG{3}))=0
\end{equation}
Equivalently:
$\BG{1}(\auxx{3}(x),\,\cdot\,)
+ \BG{2}(\,\cdot\,,x) = 0 \in \Hom_{\rR}(\je \D{0}\oplus \D{2},\rR)$
for all $x \in \DG{3}$.

Again, let us first discuss the idea. Consider
$\BG{1}$ and $\BG{2}$ as maps
\begin{alignat*}{6}
\D{1} &\;\;\times\;\; && (\je \D{0} \oplus \D{2}) && \to \rR\\
(\je \D{0} \oplus \D{2}) &\;\;\times\;\; && \DG{3} && \to \rR
\end{alignat*}
By
\eqref{eq:a2} and \eqref{ed3}
they descend to
\begin{alignat*}{6}
\D{1} &\; \;\times\;\; && (\je \D{0} \oplus \D{2})/\Auxx{2}(\DG{2}) && \to \rR\\
 (\je \D{0} \oplus \D{2})/\Auxx{2}(\DG{2}) & \;\;\times\;\; && \DG{3} && \to \rR
\end{alignat*}
Suppose the first pairing is nondegenerate.
We would get a map
from $\DG{2}$ to $\Hom_{\rR}((\je \D{0} \oplus \D{2})/\Auxx{2}(\DG{2}),\rR)$,
and then to $\D{1}$. This map would be $-\auxx{3}$.

Here is the complete argument.
Let $x \in \DG{3}$.
Fix any $c \in \VVpos$, and recall that $\D{1} = \cre_c(\D{0}) \oplus \DG{1}$
with $\cre_c: \D{0} \to \D{1}$ injective, by \kk{\oxds}{0}.
We have to show that there are unique
$y \in \D{0}$ and $z \in \DG{1}$ such that
$\BG{1}(\cre_cy+z,\,\cdot\,)
+ \BG{2}(\,\cdot\,,x) = 0 \in \Hom_{\rR}(\je \D{0}\oplus \D{2},\rR)$.
\begin{itemize}
\item The $\Hom_{\rR}(\je\D{0},\rR)$ part is equivalent to
\[
\BG{0}(\,\cdot\,,\cre_c y+z)
+
\BG{2}(\je \,\cdot\,,x)  = 0\;\;\in\;\;\Hom_{\rR}(\D{0},\rR)
\]
Since $z \in \DG{1}$ drops out by \eqref{ed1},
we get a unique $y \in \D{0}$ by \kk{\oxsp}{0}. 
\item The $\Hom_{\rR}(\D{2},\rR)$ part is
\[
\BG{1}(\cre_cy + z,\,\cdot\,) + \BG{2}(\,\cdot\,,x) = 0\;\;\in\;\; \Hom_{\rR}(\D{2},\rR)
\]
This condition holds automatically on\footnote{%
This is essentially the `diagonal descent' discussed before.
Explicitly, for all $a \in \DG{2}$:
\begin{align*}
\BG{1}(\cre_cy + z,a) + \BG{2}(a,x)
& = -\BG{0}(\auxx{2}(a),\cre_cy + z) + \BG{2}(a,x) && \textmd{by \eqref{eq:a2}}\\
& = -\BG{0}(\auxx{2}(a),\cre_cy) + \BG{2}(a,x) && \textmd{by \eqref{ed1}}\\
& =  \BG{2}(\je\auxx{2}(a),x)  +\BG{2}(a,x) && \textmd{by the construction of $y$}\\
& = \BG{2}(\Auxx{2}(a),x)\\
& = 0 &&\textmd{by \eqref{ed3}}
\end{align*}
}
$\DG{2}$.
Since $\D{2} = \cre_c(\DG{1}) \oplus \DG{2}$
by \kk{\oxds}{1}, the problem reduces to
$\BG{1}(\cre_cy + z,\cre_c\,\cdot\,) +
\BG{2}(\cre_c\,\cdot\,,x) = 0 \in  \Hom_{\rR}(\DG{1},\rR)$.
We get a unique $z \in \DG{1}$ by \kk{\oxsp}{1}.
\end{itemize}

\subsection{Lemma}

If $x \in \VV\VV\DERIV$ is real, $x = \swap_2 \cc{x}$, then
\begin{equation}\label{eq:a3stronger}
\BG{2}(x,(\je \auxx{3}+\mathbbm{1})(\DG{3})) = 0
\end{equation}
since $x = (x-x') + x'$
with $x-x' \in \je \D{0}\oplus \D{2}$
and $\rac{2}(x') = 0$.
Here $x'$ is obtained
by symmetrizing $x$ both in the two $V$
and in the two $\cc{V}$.

\subsection{The map $\bil{3}$}

Set $\Auxx{3} = \je \auxx{3} + \mathbbm{1}: \DG{3} \to (\VV)^{\otimes 3}\DERIV$.
Note that $\Auxx{3} = \je\km + \mathbbm{1}$ on $\IWeyl{3}$,
by \sref{binsb}.

Set $\bil{3} = \BG{3}(\Auxx{3}(\,\cdot\,),\,\cdot\,)$
on $\DG{3} \times \D{4}$.
It vanishes when we insert $\IWeyl{3}$ in the first slot
by \sref{binsb}.
To see that
$\bil{3}$ descends to $\EinG{3} \times \Ein{4}$,
we must show
$\bil{3}(\,\cdot\,,\IWeyl{4})=0$.

Let $x \in \DG{3}$, let $c \in \VVreal$, and let $y \in \D{3}$.
Then
\begin{alignat*}{6}
\LT{\bil{3}(x,\cre_cy)} & = \RT{\BG{3}(\Auxx{3}(x),\cre_cy)}\\
& = \RT{\BG{3}(\Auxx{3}(x),cy) - \BG{2}(\ann_cy,\Auxx{3}(x))}
& \CT{by \sref{klo}}\\
& = \RT{\BG{3}(\Auxx{3}(x),cy)} & \CT{by \eqref{eq:a3}}
\end{alignat*}
This calculation is used twice:
\begin{itemize}
\item We continue with $y \in \IWeyl{3}$:
\begin{alignat*}{6}
\LT{\bil{3}(x,\cre_cy)} & = \RT{\BG{3}(\Auxx{3}(x),cy)}\\
& = \RT{\BG{3}(y,c\Auxx{3}(x))} && \CT{by \sref{klp}}\\
&= \RT{- \BG{3}(\je \km(y),c\Auxx{3}(x))} && \CT{by \sref{binsb}}\\
&= \RT{- \BG{3}(\Auxx{3}(x),c\je \km(y))} && \CT{by \sref{klp}}\\
&= \RT{- \BG{2}(c\km(y),\Auxx{3}(x))} && \CT{by \sref{klo}}\\
& = \RT{0} && \textmd{by \eqref{eq:a3stronger}}
\end{alignat*}
Since $\IWeyl{4}$
is spanned by such $\cre_cy$,
we get $\bil{3}(\,\cdot\,,\IWeyl{4})=0$, as required.
\item  We continue with $y \in \DG{3}$:
\begin{alignat*}{6}
\LT{\bil{3}(x,\cre_cy)} & = \RT{\BG{3}(\Auxx{3}(x),cy)}\\
& = \RT{\BG{3}(\Auxx{3}(x),c\Auxx{3}(y)) - \BG{3}(\Auxx{3}(x),c\je \auxx{3}(y))}\\
& = \RT{\BG{3}(\Auxx{3}(x),c\Auxx{3}(y)) - \BG{2}(c\auxx{3}(y),\Auxx{3}(x))} && \CT{by \sref{klo}}\\
& = \RT{\BG{3}(\Auxx{3}(x),c\Auxx{3}(y))} && \CT{by \eqref{eq:a3stronger}}
\end{alignat*}
and we have used the fact that
$c \auxx{3}(y)$ is real.
Now \kk{\oxsp}{3} follows from
\sref{POS}
and axiom \axpos, because
$\Auxx{3}$
descends to an injective map
$\EinG{3} \to \msp{3}$
with \ff~complement by \kk{\oxsp}{2}.
\end{itemize}

\subsection{The map $\bil{4}$}
We first show that $\EinG{4} = 0$.
Pick any $c \in \VVpos$.
We have $\cre_c(\D{3}) = \D{4}$, because $\D{4}$ is `top-degree'.
Hence $\cre_c(\Ein{3}) = \Ein{4}$.
But we also have $\cre_c(\Ein{3}) = \cre_c(\EinG{3})$ by \kk{\oxds}{2}.
Hence $\Ein{4} = \cre_c(\EinG{3})$, and now
 \kk{\oxds}{3} implies 
 $\EinG{4} = 0$.

Set $\bil{4} = 0$. Condition \kk{\oxsp}{4} holds.

\subsection{The maps $\bil{k}$ with $k > 4$}
Set $\bil{k} = 0$.
Since $\EinG{k} \subset \Ein{k} = 0$, condition \kk{\oxsp}{k} holds.


\subsection{Remark: Degenerate input}

Suppose we run the algorithm with
 input $\gauge'$ that, instead of \sref{alginp}, satisfies:
\begin{itemize}
\item[\ixhermp] $\gauge'$ is $\rC$-Hermitian. Same as \ixherm.
\item[\ixkerp]
$\gauge'(\,\cdot\,,\KERN') = 0$, where $\KERN' \subset \cc{V}\DERIV$
is the $\rC$-submodule of elements $x$ with
\begin{align*}
x(\rC) & = 0\\
x(V) & = 0
\end{align*}
\item[\ixposp]
$\gauge': (\cc{V}\DERIV/\KERN')
\times (\cc{V}\DERIV/\KERN') \to \rC$ is positive definite.
\end{itemize}
Since $\KERN \subset \KERN'$, one
can think of $\gauge'$ as degenerate input.

The point that we want to make here is that
a good portion of the algorithm still works with $\gauge'$ as input:
we get \kk{\oxsp}{k}, \kk{\oxds}{k}, \kk{\oxff}{k} for all $k \leq 1$,
which includes $\bil{0},\bil{1},\EinG{0},\EinG{1},\EinG{2}$.
This can suffice for applications to general relativity.

\addtocontents{toc}{~\emph{Symmetric hyperbolicity}\par}

\section{Definition of `linear symmetric hyperbolic'}\label{sec:linsh}
In this section $\rR$ is a commutative ring that
satisfies axiom \axR~in \sref{axioms1}.
Suppose $U,T$ are $\rR$-modules, and suppose
\[
L\;\;:\;\;U \to T
\]
is an $\R$-linear map.
We
define what it means abstractly for $\LIN$
to be \emph{first order};
\emph{first order symmetric};
\emph{first order symmetric hyperbolic}.
Later we abbreviate the last two notions
to just \emph{symmetric} and \emph{symmetric hyperbolic}
for convenience.

\newcommand{\DIRAC}{\sigma}
\emph{Dirac example:}
Set $\rR = C^{\infty}(\R^4,\R)$, $U = C^{\infty}(\R^4,\C^2)$, $T = \Hom_{\rR}(U,\rR)$.
Set $\LIN(u)(a) = \RE (a^{\DAGGER}\DIRAC(u))$
for all $u,a \in U$
where
\[\DIRAC = \DIRAC^{\mu}\p_{\mu} =
\begin{pmatrix}
\p_0 + \phantom{i}\p_3 & \p_1 - i\p_2\\
\p_1 + i\p_2 & \p_0-\phantom{i}\p_3
\end{pmatrix}\]
Here $i=\sqrt{-1}$ ,
and $\p_\mu$ are the partial derivatives on $\R^4$,
and $a^{\DAGGER}$
is the conjugate transpose of $a \in U$.
Note that $L(u) = 0$ if and only if $\DIRAC(u) = 0$,
the massless Dirac / Weyl equation on Minkowski spacetime\footnote{%
This ad-hoc form the Dirac equation
suffices for the purpose of this section.
}. Example continues below.

\subsection{First order}\label{sec:firstorder}
Define $\TD{\LIN}: \rR \times U \to T$ by
\begin{equation*}
\TD{\LIN}(f,u) = \LIN(fu) - f\LIN(u)
\end{equation*}
By direct verification, the following are equivalent:
\begin{itemize}
\item For each $f \in \rR$, the map
$U \to T, u \mapsto \TD{\LIN}(f,u)$ is $\rR$-linear.
\item For each $u \in U$, the map
$\rR \to T, f \mapsto \TD{\LIN}(f,u)$
is a $T$-valued derivation\footnote{%
$\TD{\LIN}(ff',u) = f \TD{\LIN}(f',u) + f'\TD{\LIN}(f,u)$ for all
$f,f' \in \rR$.}.
\end{itemize}
If one and hence both of them hold, then we say that $\LIN$ is first order.

\emph{Dirac example continued:}
By direct calculation
$\TD{\LIN}(f,u)(a) =
\RE(a^{\DAGGER}(\DIRAC(f))u)$
for all $u,a \in U$ and $f \in \rR$.
It follows that $\LIN$ is first order.

\subsection{First order symmetric}
Assume that $\LIN$ is first order,
and that $T$ is the dual module of $U$,
\begin{equation*}
T = \Hom_{\rR}(U,\rR)
\end{equation*}
Define an $\rR$-bilinear map
$\J_{\LIN}: U \times U \to \Der(\rR)$
by
\begin{equation*}
\J_{\LIN}(a,b)(f) = \TD{\LIN}(f,b)(a)
\end{equation*}
If $\J_{\LIN}(a,b) = \J_{\LIN}(b,a)$
for all $a,b \in U$, then we say that $\LIN$ is first order symmetric.

\emph{Dirac example continued:}
$\J_{\LIN}(a,b)(f) =  \RE(a^{\DAGGER}(\DIRAC(f))b)$
for all $a,b \in U$ and all $f \in \rR$.
Since $\DIRAC(f)$ is a Hermitian matrix,
the map $\LIN$ is first order symmetric.

\subsection{First order symmetric hyperbolic}\label{sec:symhypdef}
Assume that $\LIN$ is first order symmetric.
Let $f \in \rR$. If
\begin{equation*}
\J_{\LIN}(\,\cdot\,,\,\cdot\,)(f)\;:\; U \times U \to \rR
\end{equation*}
is positive definite\footnote{%
One needs a notion of `positive definite' to make 
`first order symmetric hyperbolic' well-defined.
},
then we say that
$\LIN$ is first order symmetric hyperbolic relative to $f$,
and we call $f$ a time function for $\LIN$.

\emph{Dirac example continued:}
The operator $\LIN$ is first order symmetric hyperbolic relative to
$f\in \rR$ if and only if
$\p_0f > 0$
and
$(\p_0f)^2 > (\p_1f)^2 + (\p_2f)^2 + (\p_3f)^2$.
In this way, $\LIN$ encodes the causal structure of Minkowski spacetime.

\emph{Remark:}
If $\rR = C^{\infty}(\R^n,\R)$, $U = C^{\infty}(\R^n,\R^m)$,
$T = \Hom_{\rR}(U,\rR)$, then linear first order
symmetric, as defined in this \sref{linsh}, implies
$\LIN(u)(a) = c^{ij} a_i u_j + c^{ij\mu} a_i \p_{\mu} u_j$
and
$\J_{\LIN}(a,b)(f) =
c^{ij\mu}a_ib_j (\p_{\mu}f)$,
with $c^{ij}, c^{ij\mu} \in \rR$ and $c^{ij\mu} = c^{ji\mu}$.
Symmetric hyperbolic relative to $f$ implies
$c^{ij\mu}a_ia_j(\p_{\mu} f) > 0$ pointwise if $a \neq 0$.
This agrees with the standard definition of symmetric hyperbolicity.

The concept of symmetric hyperbolicity is due to Friedrichs \cite{FrKO}.

\section{Definition of `symmetric hyperbolic'}\label{sec:qsh}
Suppose $U,T$ are $\rR$-modules, and suppose
\begin{equation*}
\NON\;\;:\;\;U \to T
\end{equation*}
is polynomial over\footnote{%
We say that a map is polynomial over
$\R$ if it is a finite sum of
symmetric
$\R$-multilinear maps $U \times \cdots \times U \to T$
evaluated on the diagonal.
}
$\R$.
We can then define, for every $u \in U$,
the directional derivative
$\DNON[u] : U \to T$ by
\[
\DNON[u](b) = \tfrac{\dd}{\dd s}\big|_{s=0} \NON(u+sb)
\]
The map $\DNON[u]$ is $\R$-linear,
and it can be used in \sref{linsh}
in the role of $\LIN$.

Define: 
\begin{align*}
\text{$\NON$ is quasilinear first order} & \;\;\Longleftrightarrow\;\;
\left\{
\begin{aligned}
&\text{$\DNON[u]$ is first order for all $u\in U$}\\
&\text{$u \mapsto \TD{\DNON[u]}$ is polynomial over $\rR$}
\end{aligned}\right. \displaybreak[0]
\intertext{%
If $\NON$ is quasilinear first order,
and if $T$ is the dual of $U$, then define:}
\text{$\NON$ is symmetric} & \;\;\Longleftrightarrow\;\;
\text{$\DNON[u]$ is symmetric for all $u \in U$}
\intertext{%
If $\NON$ is symmetric,
then define for all $u\in U$ and $f \in \rR$:}\displaybreak[0]
\left.\begin{aligned}
\text{$\NON$ is symmetric hyperbolic}\\
\text{along $u$, relative to $f$}
\end{aligned}\right\} &
\;\;\Longleftrightarrow\;\;
\left\{
\begin{aligned}
&\text{$\DNON[u]$ is symmetric hyperbolic}\\
&\text{relative to $f$}
\end{aligned}\right.
\end{align*}

Quasilinear symmetric hyperbolic systems are discussed in \cite{taylor}.


\section{The linearized Einstein equations}\label{sec:lingr}

We show that the operators $\SH{k}$ defined in
\sref{shmap} are equivalent to operators
that are linear symmetric hyperbolic in the sense of \sref{linsh}.

Fix a solution $\unk' \in \Ein{1}$ to the Einstein equations,
$\eb{\unk'}{\unk'} = 0$.

Fix a $\gauge$ as in \sref{alginp}.

\subsection{Linear symmetric hyperbolicity}\label{sec:ildhdl}
The maps $\bil{k}: \EinG{k}\times \Ein{k+1}$ in \sref{algout}
induce $\rR$-module isomorphisms
\[\Ein{k+1}/\EinG{k+1} \to \Hom_{\rR}(\EinG{k},\rR)\]
by \kk{\oxsp}{k} and \kk{\oxds}{k}.
Therefore, the map $\SH{k}$ defined in \sref{shmap} is
up to an $\rR$-module isomorphism equivalent to the map
\begin{align*}
\LIN_k \;\;:\;\;\EinG{k} & \to \Hom_{\rR}(\EinG{k},\rR)\\
x & \mapsto \bil{k}(\,\cdot\,,\eb{\unk'}{x})
\end{align*}
We show that $\LIN_k$ is linear symmetric hyperbolic
in the sense of \sref{linsh}.

We use \sref{linsh}
with $\LIN = \LIN_k$ and $U = \EinG{k}$ and $T = \Hom_{\rR}(\EinG{k},\rR)$.

Identity \eqref{eq:comi} implies
\[
\J_{\LIN_k}(a,b)(f) \;=\; \bil{k}(a,\cre_{\unk'(f)}b)
\]
for all $a,b \in \EinG{k}$ and $f \in \rR$.

By \oxsp, the map $\LIN_k$ is,
in the sense of \sref{linsh},
linear symmetric hyperbolic relative
to any $f$
with $\unk'(f) \in \VVpos$.
An $f$ such that $\unk'(f) \in \VVpos$
may be called a time function for $\unk'$.

Informally, the fact that $\LIN_k$ is
symmetric hyperbolic relative to any
time function of $\unk'$ means that $\LIN_k$ propagates signals as slowly
as the causal structure of $\unk'$ allows,
that is, no faster than the speed of light.
\newcommand{\aaa}{\mathcal{Q}}
\newcommand{\ppp}{\mathcal{P}}

\subsection{Remark: Characteristic modes}

The following calculation can be informally
interpreted as saying that
$\LIN_1$ propagates only the two physical
modes at the speed of light.

For every $c \in \VVreal$ we define
the $\rR$-modules\footnote{%
One always has $\ppp(c) \subset \aaa(c)$.
If $c \in \VVpos$
then $\aaa(c) = \ppp(c) = 0$ by
\kk{\oxds}{1}.}
\begin{alignat*}{6}
\aaa(c) & \;=\; \{\;\unk \in \EinG{1}\mid \cre_c \unk \in \EinG{2}\;\}
&\qquad& \textmd{`all modes'}\\
\ppp(c) & \;=\; \{\;\unk \in \EinG{1}\mid \cre_c \unk = 0 \;\in\; \Ein{2}\;\}
&& \textmd{`physical modes'}
\end{alignat*}
We show that if $v,w \in V$ is a basis, then
\[
\aaa(v\cc{v}) = \ppp(v\cc{v})
\qquad
\textmd{and}
\qquad
\rank_{\rR} \aaa(v\cc{v}) = 2
\]

Every $\unk \in \Ein{1}$ can be expanded as
$\unk = v\cc{v}\delta' + v\cc{w}\cc{\delta} + w\cc{v}\delta + w\cc{w}\delta''$
with $\delta',\delta'' \in \Ein{0}$ and $\delta \in \DERIV$.
If we add the condition
$\unk \in \EinG{1}$, then $\delta'$ is determined by $\delta,\delta''$,
 with no restriction on $\delta,\delta''$
themselves.

For $\unk \in \aaa(v\cc{v})$
it is necessary that
$\bil{1}(\unk,\cre_{v\cc{v}}\unk) = 0$,
which implies
$\cc{v}\delta + \cc{w}\delta'' \in \KERN$,
and this condition is also sufficient for $\unk \in \aaa(v\cc{v})$.
Hence $\delta$ and $\delta''$ must annihilate
$\rC \oplus V$,
which implies that $\delta''=0$,
and $\delta$ is determined by its restriction to $\cc{V}$.
Furthermore this restriction
yields an injective $\rR$-linear map\footnote{%
Beware that, for any given $v$, different choices of
the second basis element $w$ yield
slightly different maps.
}
\begin{align*}
\aaa(v\cc{v}) & \to \Hom_{\rC}(\cc{V},\cc{V})\\
\unk & \mapsto \delta|_{\cc{V}}
\end{align*}
The image consists of the elements with vanishing trace
and image contained in $\rC \cc{v}$.
We get $\rank_{\rR}\aaa(v\cc{v}) = 2$ and
 $\aaa(v\cc{v}) \subset \ppp(v\cc{v})$, hence $\aaa(v\cc{v})
= \ppp(v\cc{v})$.


\section{The Einstein equations}\label{sec:nonlingr}

\subsection{Setup}
Fix a $\gauge$ as in \sref{alginp},
and get bilinear maps $\bil{k}: \EinG{k}\times \Ein{k+1}$
as in \sref{algout}.

Fix an $\unk_0 \in \Ein{1}$. It does not have to satisfy
the Einstein equations. In principle one can take $\unk_0$
to be equal to zero, but it is often useful to leave the choice of $\unk_0$ open. (Its role is different from the role of $\unk'$ in \sref{lingr}.)

Define the nonlinear map:
\[
\NON\;\;:\;\; \EinG{1} \to \Hom_{\rR}(\EinG{1},\rR)
\qquad \unk_1 \mapsto \tfrac{1}{2}\, \bil{1}(\,\cdot\,,\eb{\unk_0 + \unk_1}{\unk_0 + \unk_1})
\]

For all $\unk_1 \in \EinG{1}$:
\[
\NON(\unk_1) = 0
\qquad
\Longleftrightarrow
\qquad
\textmd{$\unk = \unk_0 + \unk_1$ satisfies
$\eb{\unk}{\unk} \in \EinG{2}$}
\]
Thus $\NON(\unk_1) = 0$ is a necessary condition for $\unk$
to satisfy the Einstein equations $\eb{\unk}{\unk}=0$.
We refer to $\NON(\unk_1)=0$ as the dynamical subsystem.

Observe that $\unk_1 \in \EinG{1}$
corresponds to $\unk - \unk_0 \in \EinG{1}$.
Thus the linear gauge condition for  $\unk_1$
is an affine linear gauge condition for the
original unknown $\unk$.

\subsection{Symmetric hyperbolicity of the `dynamical subsystem'}

We use \sref{qsh} with $U = \EinG{1}$
and $T = \Hom_{\rR}(\EinG{1},\rR)$.

For all $\unk_1 \in \EinG{1}$ we have
\[
\DNON[\unk_1](b) = \bil{1}(\,\cdot\,,\eb{\unk_0 + \unk_1}{b})
\]

Identity \eqref{eq:comi} implies
\[
\J_{\DNON[\unk_1]}(a,b)(f) \;=\; \bil{1}(a,\cre_{(\unk_0+\unk_1)(f)}b)
\]
for all $a,b \in \EinG{1}$ and $f \in \rR$.

By \oxsp, $\NON$ is quasilinear first order symmetric
in the sense of \sref{qsh},
and symmetric hyperbolic
along $\unk_1$
relative
to any $f$
with $(\unk_0+\unk_1)(f) \in \VVpos$.

\subsection{The `constraints propagate'}
Suppose $\unk_1 \in \EinG{1}$ solves
the `dynamical subsystem'
 $\NON(\unk_1)=0$.
We have:
\begin{itemize}
\item $C = \eb{\unk}{\unk}$ satisfies $C \in \EinG{2}$.
\item 
$\eb{\unk}{C}=0$,
by the Bianchi-Jacobi identity $\eb{\unk}{\eb{\unk}{\unk}} = 0$
in \sref{intro}.
\end{itemize}
Together they imply $C \in \ker \LIN_2$, with $\LIN_2$ from \sref{lingr},
with $\unk$ in the place of $\unk'$.

Recall that $\LIN_2$ is a linear
symmetric hyperbolic operator;
 the fact that $\unk$
is not known to solve the Einstein equations plays no role here.
Since $C \in \ker \LIN_2$ we informally conclude that
`the constraints propagate'.


\section*{Acknowledgments}
Many thanks to Horst Kn\"orrer
for commenting on drafts of this paper.
M.R.~enjoyed support from the (US) National Science Foundation.

\mostimportant{This material is based upon work supported by the National Science
Foundation under agreement No.~DMS-1128155. Any opinions,
findings and conclusions or recommendations expressed
in this material are those of the author(s) and
do not necessarily reflect the views of the National Science Foundation.}

\appendix

\addtocontents{toc}{\vskip 2mm}



\section{Recovering a conformal Riemann-Cartan geometry}\label{app:rc}
Throughout this appendix we are on a manifold $M$ as in \sref{gs}.
Recall that  $\VVreal \subset \VV$
is the set of $c$ with $c = \swap_1\cc{c}$,
and $\rank_{\rR}\VVreal = 4$.

Recall that $\IWeyl{1}=0$, and therefore
$\Ein{1} = \D{1}$.

\subsection{Definition of `nondegenerate element of $\Ein{1}$'}\label{app:NDG}
Each $\unk \in \Ein{1}$ yields,
by restriction to $\rR$,
an element of
$\VVreal
\otimes_{\rR} \Der(\rR)$.
If this element induces an $\rR$-module isomorphism
\begin{equation}\label{gind}
\Hom_{\rR}(\Der(\rR),\rR) \;\to\; \VVreal
\end{equation}
then we say that $\unk$ is nondegenerate.

The left hand side of \eqref{gind}
is (canonically isomorphic to) $\Gamma(\TCM)$, the
module of sections of the cotangent bundle.
Hence nondegeneracy entails  $\dim M = 4$.

\newcommand{\rhsrhs}{\rho}
\subsection{Lemma}\label{app:integr}
We discuss a sufficient integrability condition
for the linear equation
$\unk(a) = \rhsrhs$.
Here $\unk\in \Ein{1}$
and $\rhsrhs \in \VV\Lang'$ are given,
and $a \in \Lang'$ is the unknown.
Suppose:
\begin{itemize}
\item[(a)] $\Lang'\subset \Lang$ is a \ff~$\rC$-submodule
such that 
$\delta(\Lang') \subset \Lang'$ for all 
$\delta \in \DERIV$.
\item[(b)] $\unk \in \Ein{1}$ is nondegenerate, see \aref{NDG}.
\item[(c)] $\db{\unk}{\unk}(\rC) = 0$ and $\db{\unk}{\unk}(\Lang')=0$.
\end{itemize}

Then we have two integrability facts:
\begin{itemize}
\item (Homogeneous)
$\dim_{\C}
\{a \in \Lang'\mid\unk(a) = 0\} = \rank_{\rC}\Lang'$. If $a$
vanishes at one point of $M$ and $\unk(a)=0$, then $a=0$.
\item (Inhomogeneous)
If $\rhsrhs \in \VV \Lang'$ satisfies $\wedge_2 \unk(\rhsrhs) = 0$,
then there exists an $a \in \Lang'$ such that $\unk(a)=\rhsrhs$.
\end{itemize}

Let $(\ell_n)_{n=1,2,3,4}\subset \VVreal$ be a basis,
and expand
$\unk = \sum_n \ell_n \unk_n$
and $\rhsrhs = \sum_n \ell_n \rhsrhs_n$
with $\unk_n \in \D{0}$ and $\rhsrhs_n \in \Lang'$.
By (b) we can choose the $\ell_n$ 
so that
\begin{equation}\label{oosilmpi}
\unk_n(f) = \p_n f\qquad
\textmd{for all $f \in \rC$}
\end{equation}
with $\p_n$ the partial derivatives on $M \cong \R^4$, cf.~\sref{gs}.
The first in (c)
and $\db{\unk}{\unk}(\,\cdot\,) = 2\wedge_2 \unk(\unk(\,\cdot\,))$
imply
$\textstyle\sum_m \wedge_2\,\ell_m\unk_m(\ell_n) = 0$.
Hence for all $b \in \Lang'$,
\begin{alignat*}{6}
\unk_m(\unk_n(b)) & = \unk_n(\unk_m(b))
&\qquad&\text{by the second in (c)}\\
\unk_m(\rhsrhs_n) & = \unk_n(\rhsrhs_m)
&&\text{by $\wedge_2 \unk(\rhsrhs)=0$}
\end{alignat*}
Therefore $e_n[b] = \unk_n(b)-\rhsrhs_n \in \Lang'$ satisfies
\begin{equation}\label{eq:uiis}
\unk_m(e_n[b]) = \unk_n(e_m[b])
\end{equation}
Pick any value for $a$ at the origin of $M \cong \R^4$.
Construct $a$ along the 1-axis by solving
$e_1[a]=0$;
this is a linear ordinary differential equation along the 1-axis
by \eqref{oosilmpi} that can be solved by (a).
Construct $a$ in the 12-plane by solving
$e_2[a]=0$,
and use \eqref{eq:uiis} to conclude that $e_1[a]=0$ holds in the
12-plane because it holds on the 1-axis.
Construct $a$ in the 123-plane by solving $e_3[a]=0$,
and use \eqref{eq:uiis} again, and so forth.
The system $e_{1,2,3,4}[a]=0$ is equivalent to $\unk(a)=\rhsrhs$.

\subsection{Graded Lie algebra associated to any $\rR$-module}
\newcommand{\MDL}{E}
To every $\rR$-module $\MDL$ we can associate
\[
\GLS{\MDL} \;\;=\;\; \textstyle\bigoplus_{k \geq 0}
\; (\wedge^k \MDL) \otimes_{\rR}
\Big(\begin{array}{c}
\textmd{graded derivations on the}\\
\textmd{free $\rR$-algebra generated by $E$}
\end{array}\Big)
\]
It is a real graded Lie algebra with bracket
\begin{equation*}
\wbx{x}{y}{\MDL}
\;=\;
{\wedge_{k+\ell}} \circ {x} \circ {y} - (-1)^{k\ell}
{\wedge_{k+\ell}} \circ {y} \circ {x}
\end{equation*}
with $k,\ell$ the degrees of $x,y$,
and notation analogous to \sref{gliea}.
Note that
every module isomorphism $\MDL\to \MDL'$ induces an isomorphism
$\GLS{\MDL}\to \GLS{\MDL'}$.

\subsection{Conformal Riemann-Cartan geometry\\
associated to a nondegenerate $\unk \in\Ein{1}$}\label{app:metricricci}

A nondegenerate $\unk\in \Ein{1}$ defines
a graded Lie algebra homomorphism $h_{\unk}$ by:
\[
\xymatrix{
\D{}  \ar[rrrr]^{\textmd{canonical homomorphism}\qquad}\ar[rrrrd]_{h_\unk} &&&&
\GLS{\VVreal} \ar[d]^{\textmd{isomorphism
induced by $\unk$ through \eqref{gind}}}\\
&&&& \GLS{\Gamma(\TCM)}
}
\]

\newcommand{\guom}{g_{\unk,\omega}}
\newcommand{\nabu}{\nabla_{\hskip-2pt\unk}}
Associated to $\unk$ are a conformal metric and an affine connection:
\begin{itemize}
\item
For every one-element basis $\omega \in V\wedge V$
take $-\omega \cc{\omega}$ and permute
the four factors to get a symmetric element of $\VVreal\otimes_{\rR} \VVreal$,
which via \eqref{gind} yields a
$\guom \in \Gamma(\TCM) \otimes_{\rR} \Gamma(\TCM)$.
It is a metric with signature $\SIG$, cf.~\sref{sigfunc}.
The set $\{\guom\}_{\omega}$
ranging over all $\omega$
is a conformal metric.
\item Note that an element of degree one of
$\GLS{\Gamma(\TCM)}$ is
an affine connection
if and only if 
its restriction to $\rR$
is the differential $\dd: \rR \to \Gamma(\TCM)$.
By construction, $\nabu = h_\unk(\unk)$ is an affine connection.
\end{itemize}
For all $\unk,\omega$ there is an $\eta \in \VV$
such that $\unk(\omega) = \eta \omega$,
hence 
$\unk(-\omega\cc{\omega})
= \theta (-\omega\cc{\omega})$
with $\theta
=
\eta + \swap_1 \cc{\eta} \in \VVreal$,
hence
$\nabu \guom = \text{(1-form)}
\otimes \guom$.

The map
$\unk\mapsto(\{\guom\}_{\omega},\nabu)$
takes nondegenerate
elements of $\Ein{1}$ to conformal Riemann-Cartan geometries,
as in \sref{crccrc}.


\subsection{Solution to the Einstein equations}
Suppose a nondegenerate $\unk\in \Ein{1}$
satisfies  $\eb{\unk}{\unk}=0$.
Recall the definition of $\IWeyl{2}$ in \sref{IWeyl2},
and note that
$h_\unk(\IWeyl{2}) \ni h_\unk(\db{\unk}{\unk}) = \wbx{\nabu}{\nabu}{\Gamma(\TCM)}$. Note that:
\begin{itemize}
\item We have $\db{\unk}{\unk}(\rC)=0$,
therefore $\wbx{\nabu}{\nabu}{\Gamma(\TCM)}(\rR)=0$,
which means that the torsion of $\nabu$ vanishes.
\item 
We have $\db{\unk}{\unk}(\Lang') = 0$ where $\Lang' = V \wedge V$,
and \aref{integr} yields
$\unk(\omega) = 0$ for some one-element basis $\omega \in \Lang'$.
Then $\nabu \guom = 0$ for this particular $\omega$.
\end{itemize}
Hence $\nabu$ is the unique Levi-Civita connection
of this metric $\guom$.
The Riemann curvature
is the restriction of
$\wbx{\nabu}{\nabu}{\Gamma(\TCM)}$ to $\Gamma(\TCM)$,
and its Ricci curvature vanishes by the definition of $\IWeyl{2}$.


\end{document}